\documentclass[11pt, letterpaper]{arxiv}
\usepackage[all]{hypcap}
\usepackage[comma,numbers,sort,compress]{natbib}
\usepackage{hyperref}
\usepackage[capitalize,noabbrev]{cleveref}

\hypersetup{
 colorlinks = true,
 citecolor = {CalGoldHex},
 urlcolor={BerkeleyBlue},
 linkcolor={CalGoldHex}
}

\usepackage{microtype}

\usepackage{subcaption}
\usepackage{booktabs} %

\usepackage{algorithm}
\usepackage{algorithmic}
\usepackage{mathrsfs}
\usepackage{setspace}

\usepackage{multicol}
\usepackage{multirow}

\usepackage{amsmath}
\usepackage{amssymb}
\usepackage{mathtools}
\usepackage{amsthm}

\usepackage[most,skins,theorems]{tcolorbox}
\tcbset{
  aibox/.style={
 width=\linewidth,
 top=7pt,
 bottom=2pt,
 colback=blue!6!white,
 colframe=black,
 colbacktitle=black,
 enhanced,
 center,
 attach boxed title to top left={yshift=-0.1in,xshift=0.15in},
 boxed title style={boxrule=0pt,colframe=white,},
  }
}
\newtcolorbox{AIbox}[2][]{aibox,title=#2,#1}

\usepackage{wrapfig}
\definecolor{lightblue}{rgb}{0.22,0.45,0.70}%
\usepackage{tabularx}

\definecolor{rliableolive}{HTML}{BBCC33}
\definecolor{rliableblue}{HTML}{77AADD}
\definecolor{rliablered}{HTML}{EE8866}

\usepackage{crossreftools}
\usepackage[suppress]{color-edits}

\usepackage{dsfont}
\usepackage{nicefrac}
\usepackage{tabularx}
\newtcolorbox{analysisbox}[1][]{
 enhanced jigsaw,
 colback=white,
 colframe=blue!75!black,
 fonttitle=\bfseries,
 boxsep=5pt,
 left=5pt,
 right=5pt,
 top=5pt,
 bottom=5pt,
 title=#1,
}
\definecolor{editInitialResponse}{RGB}{255, 235, 156} %
\definecolor{editBacktrack}{RGB}{0, 0, 139} %
\definecolor{editRevisedResponse}{RGB}{255, 182, 193} %

\usepackage{pifont}
\usepackage{soul}
\definecolor{highlightmistake}{RGB}{255, 179, 179} 
\definecolor{highlightcorrect}{RGB}{179, 255, 179}

\usepackage[capitalize,noabbrev]{cleveref}

\theoremstyle{plain}

\theoremstyle{definition}

\theoremstyle{remark}

\usepackage[textsize=tiny]{todonotes}

\usepackage{enumitem}

\usepackage{amsmath,amsfonts,bm}

\def\eqref#1{Eq.~\ref{#1}}

\def\1{\bm{1}}

\DeclareMathAlphabet{\mathsfit}{\encodingdefault}{\sfdefault}{m}{sl}
\SetMathAlphabet{\mathsfit}{bold}{\encodingdefault}{\sfdefault}{bx}{n}

\title{A Theoretical Framework for Graph-based Digital Twins for Supply Chain Management and Optimization}

\author[1,2$\star$]{Azmine Toushik Wasi}
\author[1,2$\star$]{Mahfuz Ahmed Anik}
\author[1,2]{Abdur Rahman}
\author[1,2]{Md. Iqramul Hoque}
\author[1,3]{MD Shafikul Islam}
\author[1,4$\dagger$]{Md Manjurul Ahsan}

\affil[1]{Computational Intelligence and Operations Laboratory, Bangladesh}
\affil[2]{Shahjalal University of Science and Technology, Sylhet, Bangladesh}
\affil[3]{Lousiana State University, Baton Rouge, LA 70803, USA}
\affil[4]{University of Oklahoma, Norman, OK 73019, USA}
\affil[$\star$]{Equal Contribution}
\affil[$\dagger$]{Corresponding Author (\href{ahsan@ou.edu}{\texttt{ahsan@ou.edu}})}

\begin{document}

\begin{abstract}

\vspace{-0.05cm}
\textbf{Abstract:}  Supply chain management is growing increasingly complex due to globalization, evolving market demands, and sustainability pressures, yet traditional systems struggle with fragmented data and limited analytical capabilities. Graph-based modeling offers a powerful way to capture the intricate relationships within supply chains, while Digital Twins (DTs) enable real-time monitoring and dynamic simulations. However, current implementations often face challenges related to scalability, data integration, and the lack of sustainability-focused metrics. To address these gaps, we propose a Graph-Based Digital Twin Framework for Supply Chain Optimization, which combines graph modeling with DT architecture to create a dynamic, real-time representation of supply networks. Our framework integrates a Data Integration Layer to harmonize disparate sources, a Graph Construction Module to model complex dependencies, and a Simulation and Analysis Engine for scalable optimization. Importantly, we embed sustainability metrics—such as carbon footprints and resource utilization—into operational dashboards to drive eco-efficiency. By leveraging the synergy between graph-based modeling and DTs, our approach enhances scalability, improves decision-making, and enables organizations to proactively manage disruptions, cut costs, and transition toward greener, more resilient supply chains.

\end{abstract}
\maketitle

\abscontent

\section{Introduction}
In today’s world of rapid globalization and technological progress, supply chains have become highly complex and interconnected, spanning multiple continents and involving various stakeholders such as manufacturers, suppliers, distributors, and customers. These stakeholders generate vast amounts of real-time data, and managing this influx is becoming increasingly difficult due to volatile market conditions, changing consumer preferences, and large-scale disruptions like natural disasters or pandemics. Organizations must navigate these challenges while striving to maintain competitive pricing, ensure timely deliveries, and build operational resilience, all under significant time and cost pressures \citep{mark_holmes_2024}. However, modern supply chain management is not just about coordinating logistics—it involves managing intricate interdependencies across multi-level networks, where a single disruption, such as a labor strike or port congestion, can trigger widespread instability across the entire system \citep{anssi_kki__2015,Kumar2024,Taj2023aaa,Shoomal2024}. Traditional supply chain models, which rely on linear or isolated approaches and static historical data, often fail to capture the dynamic and interconnected nature of today’s global networks \citep{william_b__haskell__2024}. While these conventional methods may support basic forecasting or retrospective analysis, they struggle to address the real-time complexities and cascading effects of modern supply chains \citep{rahul_c__basole__2016}. As a result, many organizations remain reactive, responding to disruptions only after they have caused significant damage, underscoring the urgent need for more dynamic and integrated solutions in supply chain management.

\begin{figure}
    \centering
    \includegraphics[width=\linewidth]{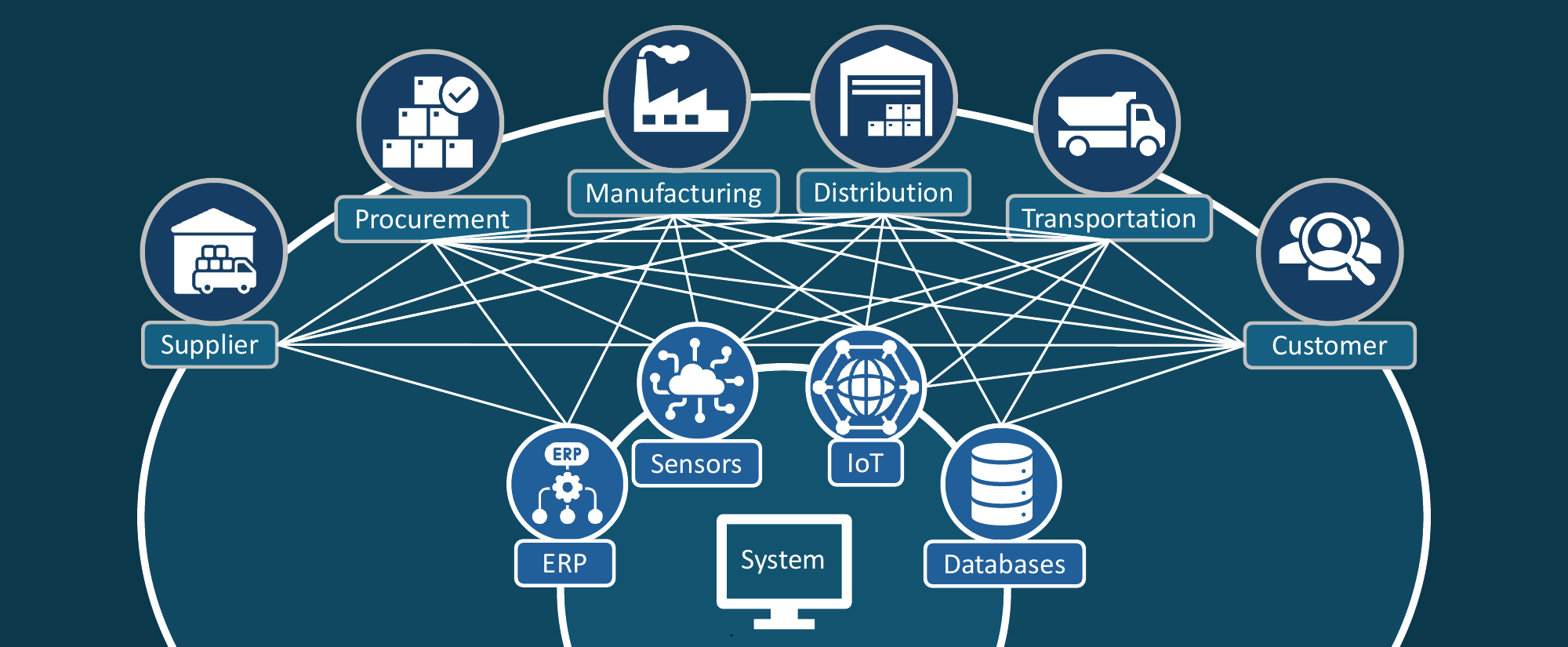}
        \caption{Different types of relations in Supply Chain}
    \label{fig:1_relations}
\end{figure}

Graph-based modeling has emerged as a powerful approach to tackling the complexities of modern industrial setups \citep{Sejan2024,Shen2023}. By representing supply chains as networks, this method uses nodes to signify key entities—such as suppliers, factories, ports, and distribution centers—and edges to illustrate the connections between them, including product movements, financial transactions, and data exchanges \citep{ting_dong__2024}, as illustrated in Figure \ref{fig:1_relations}. Unlike traditional linear models, graphs offer the flexibility to capture intricate relationships, multi-layered dependencies, and hierarchical structures that define real-world supply networks \citep{the_nonlinear}. This capability is particularly important for modeling multi-tiered supplier-customer interactions, where disruptions at one level can cascade throughout the entire system. Additionally, graph-based models help supply chain managers quickly identify vulnerabilities, such as critical nodes that act as bottlenecks or single points of failure \citep{ting_dong__2024}. The interconnected structure of graphs also makes it possible to simulate the impact of disruptions—such as delays in raw material shipments—and assess their ripple effects on inventory, production schedules, and customer satisfaction \citep{dongni_hu__2023}. By providing a more comprehensive and dynamic view of supply chains, graph-based modeling enables better decision-making and risk management in an increasingly uncertain global landscape.

\begin{figure}
    \centering
    \includegraphics[width=\linewidth]{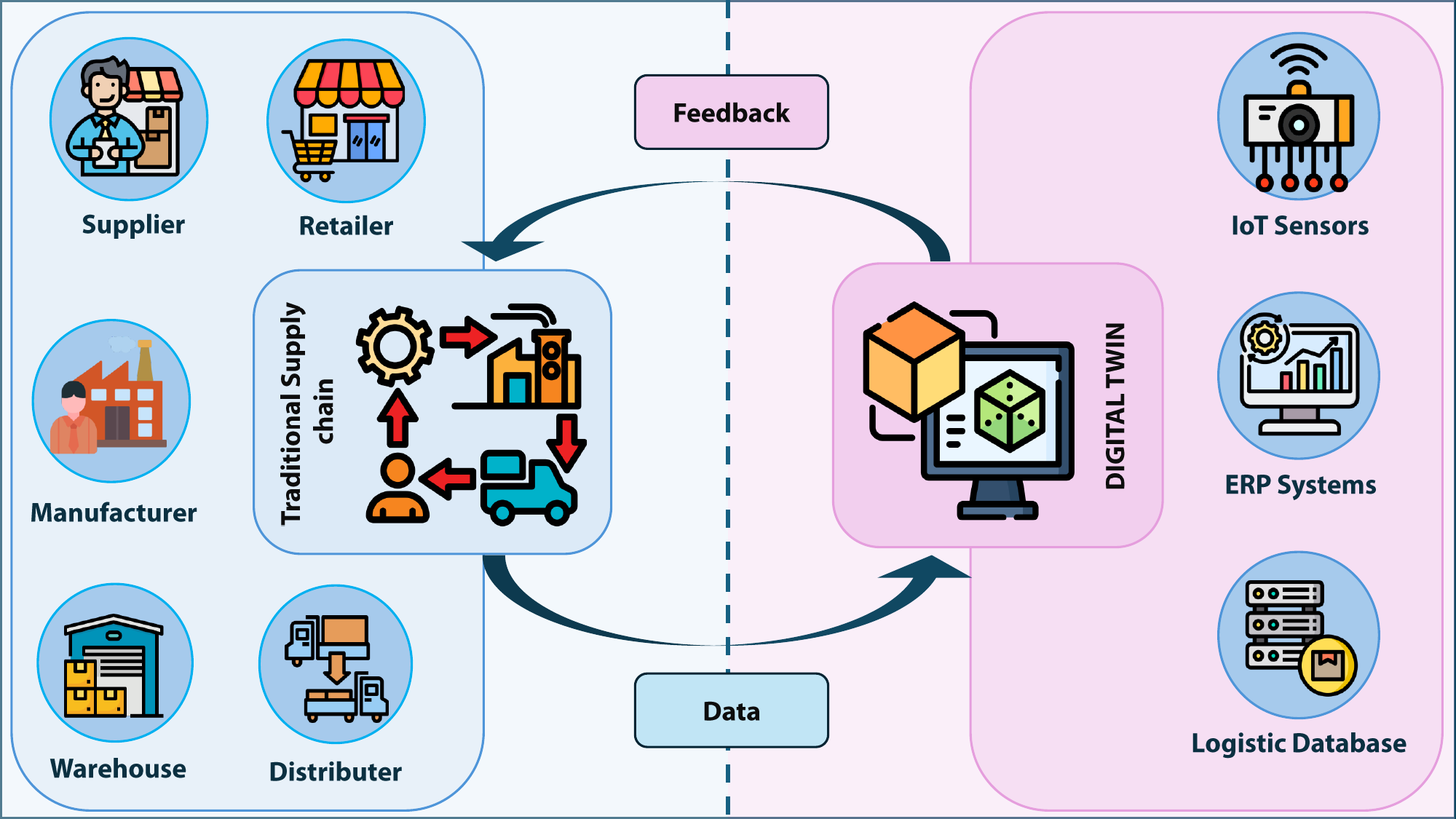}
    \caption{Digital Twins in Supply Chain Management}
    \label{fig:1_Introduction}
\end{figure}

Digital Twins (DTs), originally developed by NASA to simulate spacecraft systems \citep{allen2021digital}, have become a key technology in Industry 4.0, widely used for modeling, monitoring, and optimizing physical assets \citep{singh2021digital,jin2024big,Burattini2025}. A Digital Twin is essentially a virtual replica of a real-world entity—ranging from individual machines to entire supply chains—that enables continuous data exchange between the physical and digital environments. This bi-directional flow of information allows for real-time updates and closed-loop feedback, ensuring that changes in one domain are immediately reflected in the other \citep{abhinav_parashar_a_singh__2024,Fornari2025}. Over time, DTs have evolved from basic “digital shadows,” where data flows only one way, into fully synchronized systems capable of real-time analytics, predictive simulations, and autonomous decision-making \citep{frick2024design,Fornari2025,Burattini2025}, as demostrated in Figure \ref{fig:1_Introduction}. Industries such as aerospace, healthcare, automotive, and smart city management have already embraced DTs for predictive maintenance and scenario planning \citep{m__mythily__2024}, and their impact on supply chain management is increasingly significant. DTs provide unmatched visibility into supply chain operations, tracking material flows, inventory levels, production schedules, and transportation routes using real-time data from IoT, blockchain, and cloud computing \citep{adeoluwa_omoyemi_yekeen__2024}. This allows organizations to proactively manage risks by forecasting disruptions such as port congestion or inventory shortages \citep{ezekiel_onyekachukwu_udeh__2024}. Additionally, DTs enhance predictive and prescriptive analytics, where live data simulations help anticipate and mitigate the effects of disruptions—such as delayed shipments or production halts—while also suggesting optimal response strategies to minimize downtime \citep{mark_holmes_2024}. However, despite their advantages, implementing DTs in supply chains comes with challenges, including integrating high-quality data, ensuring security, and managing the complexity of global networks, which demand scalable architectures and advanced analytics \citep{tarun_kumar_vashishth__2024,manuel_enrique__2024}. A promising solution to these challenges is the integration of Graph Neural Networks (GNNs) with DTs. GNNs, which excel at processing graph-structured data, can help model intricate supply chain networks, detect vulnerabilities, and forecast disruptions in real time \citep{ahn2024gnn,fabian_raul_gonzalez_2024}. By combining DTs with GNN-driven insights, organizations can develop more resilient and data-driven supply chains, ensuring efficiency and adaptability in an increasingly complex and uncertain global landscape.

Integrating graph-based models into Digital Twins (DTs) enhances their ability to dynamically represent entire supply chains in real time \citep{wasi}. As data continuously flows in from sensors, market trends, or production updates, the graph structure within the DT adjusts, refining predictions and enabling swift responses to potential risks. This synergy is further strengthened by advanced machine learning techniques like Graph Neural Networks (GNNs), which effectively analyze both the structural connections within the network and individual node attributes, making them particularly useful for complex supply chain data \citep{ahn2024gnn}. By learning from historical patterns and adapting to real-time changes, GNN-powered DTs can anticipate disruptions such as production slowdowns, shipping delays, or sudden demand fluctuations \citep{fabian_raul_gonzalez_2024}. Additionally, emerging research highlights that integrating blockchain technology with graph-based DTs can improve data security and traceability, which is especially critical in highly regulated industries like pharmaceuticals and aerospace, where product authenticity and compliance are essential \citep{liang_zhou_2024,rhea_m_thomas__2024}. However, to fully leverage these advancements, standardized interoperability and strong data governance frameworks must be established, ensuring smooth communication and collaboration across all stakeholders in the supply chain.

Modern supply chains face several ongoing challenges that impact efficiency and adaptability. One major issue is managing data, as many organizations rely on disconnected information systems that were not designed for seamless integration. This fragmentation makes it difficult to create a unified view of supply chain operations, slowing decision-making and reducing transparency. Another challenge is the shortage of skilled professionals with expertise in operations research, data engineering, and machine learning, which are essential for developing and maintaining advanced Digital Twin (DT) solutions. On the technical side, scalability remains a significant hurdle. Expanding graph-based DTs to model global supply chains requires powerful computing infrastructure and efficient algorithms capable of handling vast networks with thousands of interconnected nodes and real-time data updates \citep{joshua_c__nwokeji__2019}. Additionally, as companies shift towards more sustainable supply chains, they face growing regulatory and public pressure to reduce carbon footprints, optimize energy use, and minimize waste. However, many existing systems lack the necessary tools to track and improve sustainability metrics in real-time, further complicating the transition to greener operations \citep{blessing_ameh_2024}.

\begin{figure}
    \centering
    \includegraphics[width=\linewidth]{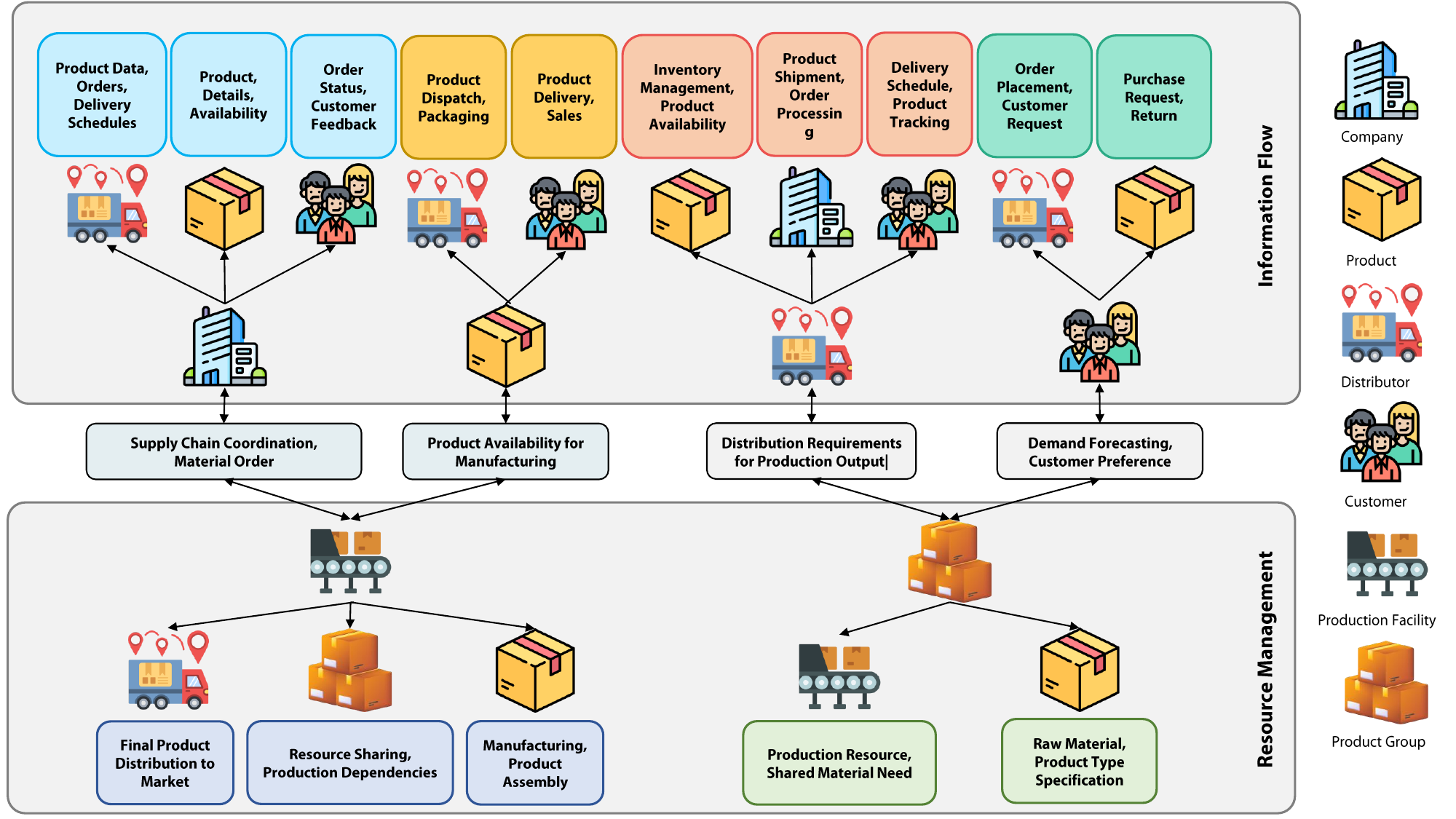}
    \caption{Different Types of Entities in Supply Chain Management and Their Dependencies}
    \label{fig:2_SCM_Connectivity}
\end{figure}

In this work, we propose a Graph-Based Digital Twin Framework for Supply Chain Optimization to address key challenges in modern supply chain management. Our framework combines advanced graph-based modeling with digital twin architecture to overcome data fragmentation, scalability issues, and sustainability concerns. We design a Data Integration Layer to seamlessly collect and harmonize information from various sources, enabling a unified and dynamic representation of supply chain networks. Using this data, we construct graph-based models that capture complex relationships and dependencies, allowing for real-time simulations and optimizations through a Simulation and Analysis Engine capable of handling large-scale networks. Additionally, we incorporate eco-efficiency metrics directly into operational dashboards, empowering organizations to track resource usage and environmental impact in real time. By leveraging the structural insights of graphs alongside real-time connectivity, our approach enhances decision-making, identifies inefficiencies, and predicts potential disruptions with greater accuracy. While further research and industry-wide standardization are necessary to address remaining challenges, the potential benefits—such as reduced downtime, cost savings, and more sustainable logistics—make this an exciting direction for the future. As the field advances, we believe that widespread adoption of graph-based digital twins could fundamentally transform how supply chains are designed, managed, and optimized in an increasingly dynamic and competitive global market, as shown in Figure \ref{fig:2_SCM_Connectivity}.

\begin{AIbox}{Our core contributions are summarized as below:} 
\begin{enumerate}
   \item We introduce a Graph-Based Digital Twin Framework, combining advanced graph modeling and digital twin technology to address challenges in data integration, scalability, and optimization within supply chain management.
    \item Our framework includes a Graph Construction Module and a Simulation and Analysis Engine, enabling real-time modeling and optimization of complex and large-scale supply networks.
    \item We integrate eco-efficiency metrics into dashboards, helping organizations track and improve sustainability aspects like carbon emissions, energy use, and waste reduction in real-time.
    \item Our framework provides a constantly updated and clear view of the supply chain using graph-based connectivities, helping to identify inefficiencies, predict problems, and make faster, informed decisions.
\end{enumerate}
\end{AIbox}

In the following sections, we provide a detailed exploration of the various aspects that underpin our work. Section \ref{sec:Background} introduces the foundational concepts of graphs, supply chains, and digital twins, providing the necessary background for understanding how these elements come together in our proposed framework. In Section \ref{sec:RelatedWorks}, we review existing literature and related works that inform and contextualize our approach, highlighting the gaps that our research aims to address. Section \ref{sec:motvation} outlines the key motivations behind this work, discussing the challenges within supply chain management and how our framework can help overcome them. Next, in Section \ref{sec:archi-emthodology}, we describe the architecture and methodology of our Graph-Based Digital Twin Framework, detailing its components and how they work together to optimize supply chains. Section \ref{sec:discussion} offers an in-depth discussion on the implications of our framework, exploring potential use cases, benefits, and limitations. In Section \ref{sec:Challenges-and-research-directions}, we identify the current challenges in the field and suggest future research directions that could further enhance the capabilities of digital twins in supply chain management. Finally, Section \ref{sec:conclusion} wraps up the paper, summarizing our findings and offering concluding remarks on the potential impact of our work.

\section{Background and Fundamentals} \label{sec:Background}
In this section, we will explore the key concepts behind modern supply chain optimization. We’ll look at how graph-based models can help us understand the connections within a supply chain, how dynamic graphs can track changes over time, and how Graph Neural Networks (GNNs) can improve predictions and decision-making. Together, these tools help us better manage and improve the complex systems that drive supply chains.

\subsection{Supply Chain and Graphs}
\subsubsection{Supply Chain as Nodes \& Edges}
In our approach to modeling supply chains, we use graphs to represent the interconnected relationships between different entities. A graph consists of nodes and edges: nodes represent the key players in the supply chain, while edges capture the connections and interactions between them. In supply chain management (SCM), these nodes include suppliers, manufacturers, distributors, retailers, consumers, and logistics providers. Each of these entities has specific characteristics that define their role in the supply chain. For example, suppliers deliver raw materials and are defined by attributes like production capacity, location, and lead times \citep{Hu2022}. Manufacturers, who turn raw materials into finished goods, are characterized by factors such as production costs and capacity \citep{1}. Distributors manage inventory and handle the flow of goods from manufacturers to retailers, with attributes like inventory turnover rates affecting their operations \citep{pratik_j__parikh__2010}. Retailers directly engage with consumers, maintaining stock levels based on demand patterns \citep{shikhar_srivastava__2023}. Logistics providers, including third-party and fourth-party logistics (3PL/4PL), ensure the efficient transportation and delivery of goods, focusing on cost optimization and reliability \citep{shreya_andru_2024}.

The edges in this graph represent the relationships and interactions between these entities. These include transactions, transportation routes, information flows, and collaborations. Each edge has properties that provide insights into how these relationships function, such as weight, directionality, and temporal attributes. For instance, transportation routes between suppliers and manufacturers are defined by factors like cost, transit time, and environmental impact \citep{jyri_vilko__2012}, all of which affect efficiency and sustainability. Transactions between entities—whether involving materials, finances, or services—are characterized by parameters such as value, volume, and delivery schedules, which are crucial for maintaining the economic flow of the supply chain \citep{zillur_rahman__2007}. Information flows are key for synchronizing the supply chain, as they allow for the exchange of data on demand forecasts, inventory levels, and production schedules, helping to reduce inefficiencies \citep{Ahmad2023}. Collaborations, whether through joint ventures or shared resources, rely on trust, mutual goals, and risk-sharing, which create synergies and competitive advantages \citep{Lyasko2020}. Finally, edges can also represent dependencies, such as reliance on a single supplier, which can reveal vulnerabilities in the supply chain and highlight the need for risk mitigation strategies.

By using a graph-based approach, we can model the complex web of relationships that make up modern supply chains. This framework helps us visualize and analyze the flow of goods, services, and information across the network, identify potential bottlenecks, and optimize operations in real-time. Moreover, since supply chain relationships are constantly evolving, graph-based models are particularly useful for managing dynamic systems. With this approach, we can not only improve operational efficiency but also gain a deeper understanding of the risks and vulnerabilities present in these complex networks.

\subsubsection{Dynamic Graphs in Supply Chain}
Dynamic graphs offer a powerful way to model supply chain networks by capturing the ever-changing nature of relationships and operations. Unlike static graphs, which show fixed connections and attributes, dynamic graphs account for changes over time, such as shifts in inventory levels, production rates, and disruptions across the network \citep{mayer1995local}. This time-sensitive approach is crucial for modern supply chains, where factors like fluctuating demand, unexpected disruptions, and varying transportation lead times require constant adjustment. For instance, dynamic graphs can simulate how disruptions ripple through the network, helping to identify vulnerable points or critical paths that need immediate attention \citep{Ji2024}. They also support real-time inventory management, triggering alerts when stock levels fall below critical thresholds and suggesting emergency replenishment from nearby warehouses or retailers, considering factors like transit times and costs \citep{Chartniyom2007OptimalIR}. Additionally, dynamic graphs enable the tracking of production rates, allowing managers to identify bottlenecks or overproduction in specific facilities and make timely adjustments to resources \citep{krishna_k__krishnan__2006}. Overall, this approach improves supply chain responsiveness and enhances operations by continuously adapting to real-time conditions.

\subsubsection{Defining Supply Chain as a Dynamic Graphs}
We model a dynamic supply chain network as a time-varying graph $G(t) = (V, E(t))$, where $V$ denotes the set of nodes representing supply chain entities (such as suppliers, manufacturers, distributors, retailers, etc.), and $E(t)$ represents the set of edges at time $t$ that capture the relationships between these entities. Although the set $V$ may remain constant or vary with time, the key aspect is that the attributes associated with both nodes and edges evolve dynamically.

\begin{tcolorbox}[colback=green!5!white,colframe=black,boxsep=0pt,top=4pt,bottom=4pt,left=3pt,right=3pt]
For each node $i \in V$, we associate a state vector $x_i(t) \in \mathbb{R}^n$ at time $t$ that encapsulates its operational attributes (for example, inventory levels, production capacity, or lead times). The evolution of each node’s state is modeled by a discrete-time dynamic equation:
\begin{equation}
x_i(t+1) = f_i\Bigl(x_i(t),\, u_i(t),\, \sum_{j \in N_i(t)} \phi_{ij}\bigl(w_{ij}(t),\, x_j(t)\bigr),\, \xi_i(t)\Bigr),
\end{equation}
where $u_i(t)$ represents control actions applied at node $i$ (such as ordering or production decisions), $N_i(t)$ is the set of neighboring nodes connected to $i$ at time $t$, $\phi_{ij}$ quantifies the influence from node $j$ to node $i$ through the edge $(i,j)$ with weight $w_{ij}(t)$, and $\xi_i(t)$ captures any external disturbances or noise affecting node $i$.

Similarly, the edges $(i,j) \in E(t)$ are characterized by weights $w_{ij}(t)$ that may reflect transportation costs, transit times, or capacity, and these weights are updated over time according to:
\begin{equation}
w_{ij}(t+1) = g_{ij}\bigl(w_{ij}(t),\, \eta_{ij}(t)\bigr),
\end{equation}
with $g_{ij}$ being an update function and $\eta_{ij}(t)$ representing exogenous influences or noise affecting the edge.
\end{tcolorbox}

\begin{tcolorbox}[colback=green!5!white,colframe=black,boxsep=0pt,top=4pt,bottom=4pt,left=3pt,right=3pt]
In addition to the state dynamics, we define a flow variable $f_{ij}(t)$ representing the quantity of goods or information transferred from node $i$ to node $j$ at time $t$. To ensure the conservation of flow at each node $i$, we use the following balance equation:
\begin{equation}
x_i(t+1) = x_i(t) + \sum_{j \in N_i(t)} f_{ji}(t) - \sum_{j \in N_i(t)} f_{ij}(t) + s_i(t),
\end{equation}
where $s_i(t)$ denotes any supply or production at node $i$ at time $t$.
\end{tcolorbox}

\begin{tcolorbox}[colback=red!5!white,colframe=black,boxsep=0pt,top=4pt,bottom=4pt,left=3pt,right=3pt]
To evaluate the overall performance of the supply chain over a time horizon $T$, we define a cost function that aggregates both operational and transportation costs:
\begin{equation}
J = \sum_{t=0}^{T-1} \left[ \sum_{i \in V} c_i\bigl(x_i(t),\, u_i(t)\bigr) + \sum_{(i,j) \in E(t)} c_{ij}\bigl(w_{ij}(t),\, f_{ij}(t)\bigr) \right].
\end{equation}
Our objective is to choose the control actions $\{u_i(t)\}$ and flows $\{f_{ij}(t)\}$ so as to minimize the total cost $J$, subject to the dynamic constraints on the nodes and edges.
\end{tcolorbox}

This mathematical formulation provides a rigorous framework for analyzing and optimizing supply chain operations in real time. By capturing the dynamic nature of the system through time-varying graphs, we are able to model the complex interdependencies and continuously adapt to changes in operational conditions.

\subsection{Combining Graphs and Digital Twins for Supply Chains}
Graph Neural Networks (GNNs) have revolutionized supply chain management by offering an effective way to model the complex relationships and dynamics that are inherent in supply chain networks. Unlike traditional methods, GNNs can capture non-linear interdependencies between different entities (nodes) and the connections (edges) between them. This makes them particularly useful for predictive tasks and optimization. For instance, models like Graph Convolutional Networks (GCN) and Graph Attention Networks (GAT) excel at forecasting disruptions by analyzing how different firms within the supply chain are interconnected and how disruptions—such as natural disasters or market fluctuations—might ripple through the network \citep{Su2024}. These GNN models can identify weak points in the supply chain and suggest actionable strategies for recovery, helping organizations anticipate challenges and mitigate risks before they escalate. Furthermore, GNNs are invaluable in optimizing supply chain flows, as they support real-time decision-making. They allow for dynamic adjustments to resource allocation, inventory levels, and transportation routes, all of which are essential in a fast-moving, unpredictable environment. More advanced models, such as the Hierarchical Knowledge Transferable Graph Neural Network (HKTGNN), enhance the system’s performance by analyzing the roles and interdependencies of different supply chain nodes, reducing bottlenecks, and improving operational efficiency \citep{https://doi.org/10.48550/arxiv.2311.04244}.

When we integrate GNNs with Digital Twins (DTs), the effectiveness of these models is further amplified, bringing even greater precision and adaptability to supply chain management. Digital Twins provide a real-time, digital replica of the physical supply chain, allowing GNN models to work in a highly dynamic environment where they can continually adjust based on live data. For example, when we pair Dynamic Graph Neural Networks (DGNNs) with a DT, companies gain the ability to assess policy changes, explore alternative suppliers, and make more accurate demand forecasts, especially in complex, multi-tiered supply chains \citep{https://doi.org/10.48550/arxiv.2206.03469}. The combination of GNNs and DTs creates a closed-loop system where predictions, optimizations, and simulations are constantly updated with real-world data. This dynamic interplay allows for more informed decision-making, better strategic planning, and a deeper understanding of the evolving nature of supply chains. Ultimately, this integration enhances the resilience, efficiency, and agility of supply chain operations by uncovering hidden patterns, refining forecasts, and supporting proactive adjustments in real-time.

\section{Related Works} \label{sec:RelatedWorks}
In recent years, supply chain management has seen significant advancements with the growing use of Digital Twin (DT) technologies and graph-based methodologies. These innovations have transformed how we model, analyze, and optimize complex supply chain networks. In this section, we review the existing literature on the application of Digital Twins in supply chain management, the use of graph-theoretic models for understanding supply chain dynamics, and the integration of graph-based methods within Digital Twin frameworks. By examining the current research, we aim to highlight the gaps that still exist in this field and suggest potential directions for future work. Our goal is to further develop these approaches to better tackle the increasing complexities of modern supply chains.

\subsection{Digital Twin Approaches in Supply Chain}

Digital Twin (DT) technologies have become a transformative tool in supply chain management, offering improved visibility, resilience, and efficiency by creating digital replicas of physical systems. These virtual models allow for real-time data exchange, providing actionable insights and predictive analytics that help with proactive management and ensuring operational continuity during disruptions \citep{wang2022digital}, \citep{zarnitz2022digital}. While DTs are primarily known for their ability to monitor and simulate supply chain behaviors, their applications can vary greatly depending on the industry and how they are integrated with existing technologies.
For example, \citep{moshood2021digital} explored how DTs could enhance logistics visibility by incorporating predictive metrics to optimize operational outcomes. In the pharmaceutical sector, \citep{chang2024learning} showed how DTs help mitigate ripple effects and improve supply chain agility, which is especially important in industries where disruptions can have high-stakes consequences. These examples reflect the growing recognition of DTs' role in improving both operational efficiency and responsiveness to unexpected events.

Expanding on the scalability of DTs, \citep{luo2023expanding} examined how they can optimize resource utilization and promote sustainability across multi-level supply chains. Similarly, \citep{liu2023leveraging} explored the integration of DTs in industrial symbiosis networks, where they facilitate collaboration and promote resource efficiency. These studies highlight the broad range of applications for DTs, with particular emphasis on logistics, disruption management, and sustainability.
The integration of DTs with cutting-edge technologies like artificial intelligence (AI) and machine learning has further increased their capabilities. For instance, \citep{ashraf2024disruption} proposed a Cognitive Digital Supply Chain Twin (CDSCT) framework that uses deep learning to detect disruptions and support dynamic recovery, significantly boosting system resilience. Additionally, \citep{xu2023implementation} demonstrated the use of multi-agent systems within DTs, which allow for autonomous decision-making in supply chains, leading to more robust and decentralized operations.

Real-world case studies provide further insights into the application of DTs. \citep{valero2022conceptual} introduced a circular meat supply chain DT framework that focuses on reducing waste and enhancing operational efficiency, aligning with the growing interest in sustainable supply chain practices. Similarly, \citep{zarnitz2022digital} highlighted how Digital Supply Chain Twins (DSCTs) in the automotive sector can assist in predictive planning, disruption mitigation, and sustainable logistics management.
Despite these promising developments, several challenges remain in scaling DT adoption. \citep{barykin2021place} identified technical and economic obstacles, such as issues with data interoperability, the complexity of real-time integration, and the lack of standardized frameworks, which hinder widespread implementation. Overcoming these barriers through integrated approaches and continued technological innovation will be critical to realizing the full potential of DTs in optimizing supply chains. 
In summary, the body of research on Digital Twin technologies in supply chains reveals both opportunities and challenges. While DTs have proven effective in enhancing operational efficiency, improving resilience, and enabling sustainable practices, significant hurdles remain, particularly in their widespread adoption. Addressing these challenges will be essential for fully harnessing the power of DTs in modern supply chain management.

\subsection{Graph-Centric Approaches in Supply Chain Networks}
Graph-centric methodologies have become increasingly important in improving supply chain management by taking advantage of the natural graph-like structure of supply networks. Among these methods, Graph Neural Networks (GNNs) have demonstrated exceptional performance in solving complex challenges such as demand forecasting, risk assessment, and inventory optimization, often outperforming traditional approaches. In our review, we found that studies have highlighted the ability of GNNs to model intricate relationships between supply chain entities, improving both forecasting accuracy and risk mitigation strategies \cite{wasi} and \cite{kotecha2024leveraging}. These models are especially effective in capturing both static and dynamic elements of supply chains, which is crucial for making real-time decisions. For example, \cite{chang2024learning} introduced a temporal GNN framework that accounts for the dynamic nature of supply chains, significantly improving disruption prediction and production function forecasting with an impressive accuracy increase of 62\%. Additionally, different GNN variants, such as Graph Attention Networks (GATs) and Graph Convolutional Networks (GCNs), have been explored for their superior forecasting capabilities \citep{han2024applying}, further highlighting the flexibility and robustness of GNNs in modeling both temporal and static supply chain complexities.

Furthermore, Knowledge Graphs (KGs) have played a key role in enhancing the modeling of supply chain networks, particularly in more complex, multi-tiered systems. By integrating KGs with large language models (LLMs), \cite{almahri2024enhancing} introduced a method for zero-shot learning, enabling the extraction of intricate supply chain relationships and the mapping of multi-tier networks. This approach was particularly useful in the electric vehicle supply chain, helping bridge information gaps and create more comprehensive models of supply chain dynamics. Similarly, \cite{liu2023knowledge} employed graph completion techniques to predict missing relationships between supply chain entities, which further enhanced the resilience of these networks. By mapping the connections and dependencies between different supply chain entities, KGs help organizations better understand their networks, detect vulnerabilities, and mitigate risks that could lead to disruptions. These advances underscore the vital role that graph-based approaches, particularly KGs, play in improving supply chain visibility and operational efficiency.

Lastly, graph-centric approaches have shown significant potential in risk management and inventory control, two crucial aspects of maintaining an efficient supply chain. In this area, \cite{kosasih2024towards} introduced a neurosymbolic reasoning approach that combines GNNs and KGs to uncover hidden risks within supply chains. This hybrid approach generates actionable insights that can help mitigate disruptions before they occur by understanding the complex dependencies between supply chain entities. Additionally, the integration of GNNs with Multi-Agent Reinforcement Learning (MARL) has been explored for decentralized inventory control \citep{kotecha2024leveraging}. This combination allows for more adaptive and responsive inventory management, which is particularly useful in uncertain and volatile environments. These applications highlight how graph-based methodologies can significantly enhance the predictive accuracy, streamline operations, and strengthen the overall resilience of supply chains, offering deeper insights and more agile, data-driven decision-making processes.

\subsection{Research Gap Analysis and Our Contribution}
Despite the growing interest in Digital Twin (DT) technologies and graph-based approaches in supply chain management, there remains a significant gap in integrating these two paradigms to enhance supply chain operations. While DTs have been widely studied for improving supply chain visibility, resilience, and sustainability through real-time data analytics and simulation-based approaches \citep{moshood2021digital,zarnitz2022digital}, most research has focused on applying DTs in isolation. These studies often concentrate on simulating physical systems or processes without addressing the complex relationships that govern supply chain networks. On the other hand, graph-based approaches, especially Graph Neural Networks (GNNs), have shown promising results in optimizing various supply chain functions like demand forecasting, risk management, and inventory control \citep{kosasih2024towards,wasi}. However, these approaches have not been integrated with DTs to enhance their predictive and analytical capabilities. Although some studies, such as \citep{ashraf2024disruption} and \citep{luo2023expanding}, have explored DT applications for disruption management, they have not incorporated graph-theoretical methods to model the intricate interdependencies within supply chains. This highlights a key research gap that we aim to address: the need for frameworks that combine the real-time, data-driven power of DTs with the structural insights of graph-centric models.

Moreover, the application of GNNs and other graph-based methodologies within DT systems is still in its early stages. Most current DT applications focus on simulation, disruption management, or sustainable supply chain practices \citep{valero2022conceptual,liu2023leveraging}, but they often overlook the dynamic interrelationships modeled by graph-based approaches. For instance, while \citep{xu2023implementation} explored multi-agent systems for supply chains, they did not integrate graph-theoretical concepts into their DT framework. Additionally, although GNNs have demonstrated their ability to predict hidden links \citep{aziz2021data} and manage risk propagation \citep{wasi}, their integration within DT systems for enhancing real-time predictive capabilities remains largely unexplored. The literature still lacks a unified approach that combines the strengths of DTs and graph-based models for dynamically optimizing supply chain networks. Our work addresses this gap by proposing a novel framework that integrates real-time data analytics with graph-based insights, improving supply chain decision-making in complex and interconnected environments.

In contrast to existing studies, our approach introduces a hybrid framework that combines the capabilities of DTs with graph-based methodologies, particularly GNNs, to enhance supply chain optimization and risk management. While previous research has focused on either DT-driven simulations or graph-based models separately, our approach seeks to merge these two paradigms. By incorporating GNNs into the DT framework, we aim to improve predictive accuracy, risk assessment, and network optimization, tackling both operational and structural complexities. This integrated methodology enables more effective identification of weak links, risk propagation, and dynamic network optimization, offering a more comprehensive and data-driven approach to managing disruptions and inefficiencies. Our work thus not only fills a significant gap in the literature but also advances the state of the art in supply chain management by combining the real-time, data-driven capabilities of Digital Twins with the structural insights of graph-based models.

\section{Motivation} \label{sec:motvation}
\begin{figure}
    \centering
    \includegraphics[width=\linewidth]{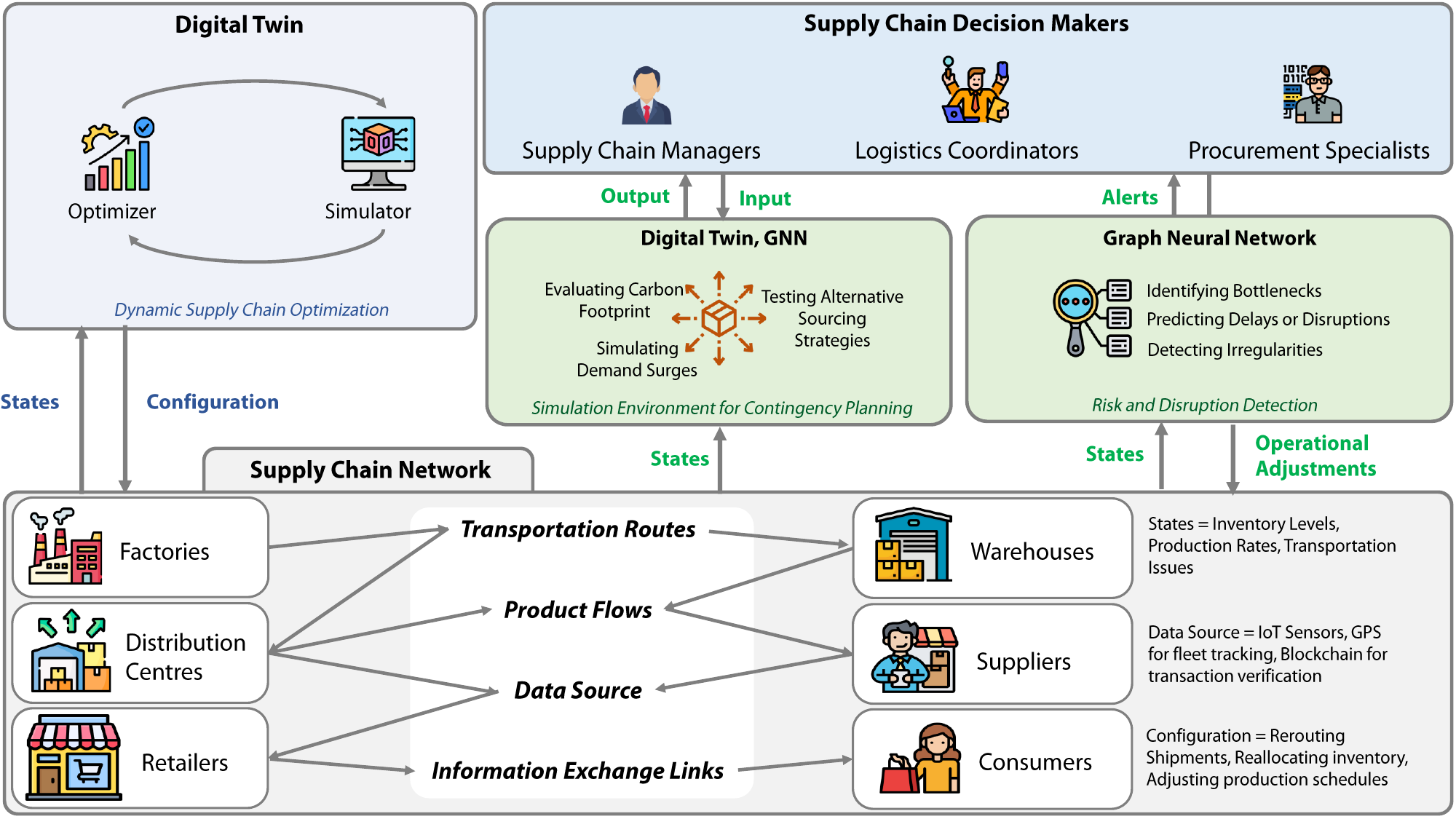}
    \caption{Motivation for Graph-based Digital Twins for Supply Chain Management and Optimization}
    \label{fig:8_Motivation_section}
\end{figure}

Our work is motivated by the critical challenges facing modern supply chain management, where traditional linear models fall short in capturing the complexities of globally dispersed, multi-tiered networks, as shown in Figure \ref{fig:8_Motivation_section}. These models often struggle to address dynamic interdependencies, real-time disruptions, and the need for adaptive decision-making in fast-paced environments. As supply chains become increasingly interconnected and vulnerable to rapid changes, there is a growing demand for innovative approaches that can offer more responsive, resilient, and data-driven solutions to effectively navigate these complexities.

\subsection{Rethinking Traditional Supply Chain Models}
Global supply chains have evolved into complex, multi-tiered networks that span across countries and involve a wide range of stakeholders, including suppliers, manufacturers, logistics providers, wholesalers, and end customers \citep{torky_althaqafi_2024}. These networks are highly distributed and interdependent, meaning that a disruption in one part of the system—whether due to a natural disaster, geopolitical tensions, or supply chain inefficiencies—can have a cascading effect across the entire system \citep{anssi_kki__2015,ashraf2024disruption}. Each component, whether it's a manufacturing plant or a transportation hub, operates under different constraints and goals, making it increasingly difficult to maintain smooth operations, especially when external factors like shifting market demands and unforeseen events come into play. Traditional supply chain models, which were designed for more stable and predictable environments, often fail to address the complex, dynamic nature of today's global supply networks \citep{taehyun_roh__2024}.

One of the key limitations of traditional models is their tendency to oversimplify supply chains, viewing them as linear, step-by-step processes with stable demand and isolated stakeholders \citep{virginia_l__m__spiegler__2015}. However, real-world supply chains are much more interconnected and prone to feedback-driven dynamics. Changes in consumer demand, for instance, can create ripple effects that disrupt upstream suppliers, while delays or errors in one part of the supply chain can cause bottlenecks in others \citep{jaeger2021identification}. This kind of complexity is often exacerbated by events like labor strikes, regulatory shifts, or port closures, which can magnify small disruptions into larger-scale problems. Static models, which rely on historical data and predefined parameters, cannot capture this volatility or provide timely insights for decision-makers, forcing companies to adopt reactive strategies rather than proactive ones.

Given these challenges, it's clear that we need to rethink the traditional supply chain models that rely on linear, static frameworks. A major gap in many of these approaches is the lack of real-time data integration, which makes it difficult to respond to emerging risks such as sudden spikes in demand or shortages of key materials \citep{abeer_aljohani_2023}. Without real-time feedback, businesses often face issues like excess inventory, production delays, or missed opportunities to reroute shipments before a disruption escalates \citep{sboui2006unsold,suganya2020determinants}. As customer expectations continue to rise, particularly in areas like rapid shipping and product customization, traditional supply chain models are increasingly inadequate. We must therefore explore new, more adaptive modeling frameworks that can better account for the interconnected and dynamic nature of modern global supply chains.

\subsection{Introducing Graph-based Digital Twins}
Digital Twins (DTs) have become a game-changer in optimizing physical assets by creating real-time connections between physical and virtual environments \citep{jin2024big}. Initially, these systems were mainly used for monitoring machines or managing product lifecycles in localized settings. However, when we try to apply them to entire supply chains, traditional DTs often fall short. They tend to rely on incomplete data and one-way information flows, which makes it hard to capture the complex, interconnected nature of global networks \citep{gary_hildebr_2024}. This is particularly challenging when trying to simulate or anticipate disruptions that can cascade across different parts of the system. In this context, graph-based modeling offers a promising solution. By treating elements like suppliers, transport routes, and distribution hubs as nodes, and their relationships (e.g., product flows or financial transactions) as edges, we can create a more dynamic and interconnected view of the supply chain \citep{ting_dong__2024}. Real-time data from sources like IoT sensors, weather reports, or geopolitical alerts further enhance this model, ensuring that supply chain managers can make proactive decisions rather than relying on outdated, static data.

Beyond just providing a better visualization, graph-based DTs open up new possibilities with advanced computational techniques designed for network-based data. Tools like centrality measures (e.g., betweenness and closeness) help identify critical nodes that could become bottlenecks during a crisis, while community detection methods can pinpoint groups of suppliers that are at risk due to shared vulnerabilities \citep{hanauer2022recent}. These features allow us to conduct "what-if" analyses to simulate the impact of various disruptions, such as a supplier failure or route congestion, and develop contingency plans in advance \citep{karras2017systematic}. Additionally, by using Graph Neural Networks (GNNs), we can analyze complex patterns in supply chains by combining data from multiple sources like shipment volumes and lead times with the underlying network structure \citep{chang2024learning}. This capability enables more accurate forecasting, identification of risks, and better demand predictions across different tiers of the supply chain. The integration of real-time data and continuous learning from the network allows GNN-powered DTs to remain adaptive, helping businesses stay agile and prepared for unexpected challenges.

However, despite their great potential, adopting graph-based Digital Twins in supply chains still comes with significant hurdles. Data integration is one of the biggest challenges, as many organizations use different IT systems, formats, and reporting methods, making it difficult to synchronize data in real time \citep{ibrahim_ahmed_albaltah__2020,doan_quang_tu__2019}. Furthermore, concerns about cybersecurity complicate the deployment of these systems, as sensitive data must be securely transmitted across complex networks \citep{wylde2022cybersecurity}. To address these issues, we propose a comprehensive framework that guides the integration of graph-based DTs in large-scale supply chains. This framework outlines key components like data ingestion pipelines, real-time analytics modules, and decision-support systems, with a focus on machine learning, particularly GNNs, to improve predictive and prescriptive capabilities. By incorporating strong data governance and security measures, along with a focus on sustainability and compliance, this framework aims to create resilient, adaptive supply chains that can respond to challenges while maintaining environmental and social responsibility. By moving from static, one-way systems to interactive, real-time models, we can build future-proof supply chain ecosystems that are prepared for the uncertainties and opportunities of an increasingly complex world.

\section{Architecture of Our Graph-based Digital Twin} \label{sec:archi-emthodology}
\begin{figure}
    \centering
    \includegraphics[width=\linewidth]{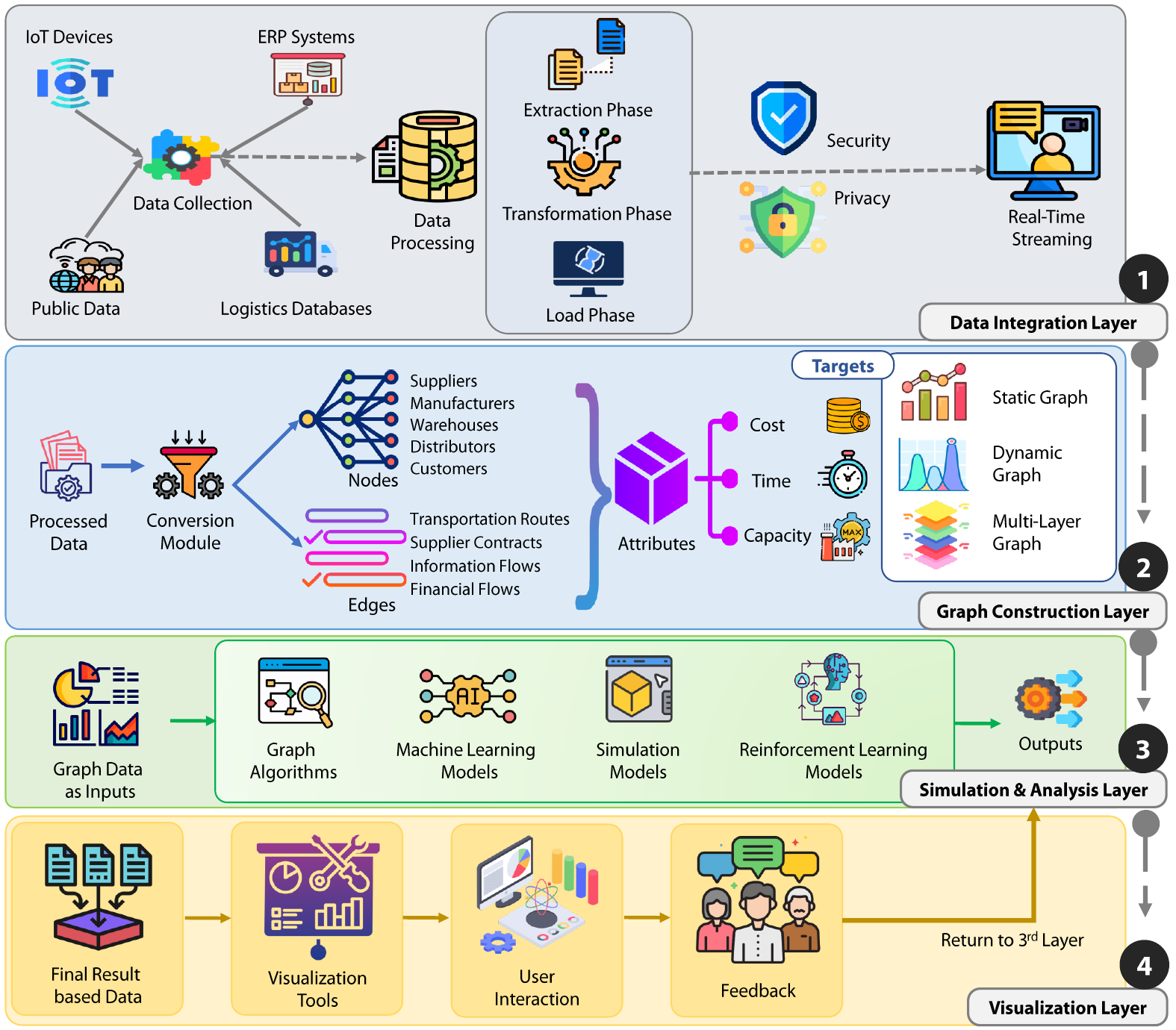}
    \caption{Our Full Architecture of Graph-based Digital Twins for Supply Chain Management and Optimization}
    \label{fig:9_Combined_methodology}
\end{figure}

In this section, we present the deatiled architecture of a Graph-Based Digital Twin (GDT) designed specifically for supply chain data. An illustration showing all the core functions and components are available in Figure \ref{fig:9_Combined_methodology}. A GDT is a dynamic, interconnected system that integrates real-time data from various sources to provide a comprehensive view of the supply chain network. The architecture consists of multiple layers, each responsible for distinct functions, ranging from data ingestion and processing to real-time analytics and decision support.

\subsection{Data Integration Layer Function}
\begin{figure}
    \centering
    \includegraphics[width=\linewidth]{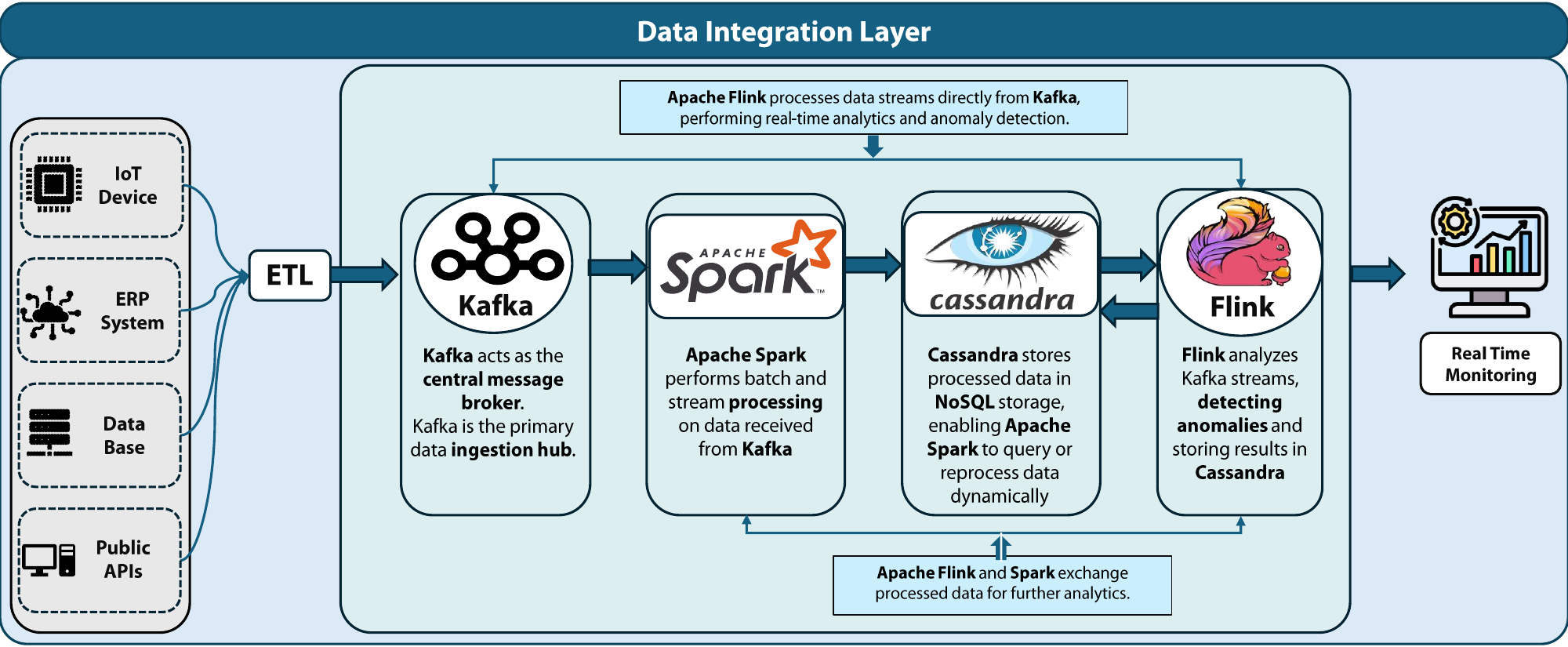}
    \caption{Workflow and Components of Data Integration Layer}
    \label{fig:3_Data_Integration_Layer}
\end{figure}

Data Integration Layer is a critical component of our Graph-Based Digital Twin (GDT) architecture, as it ensures the smooth aggregation, standardization, and preprocessing of various data sources to create an accurate and dynamic supply chain model \citep{croset2016flexible}, as illustrated in Figure \ref{fig:3_Data_Integration_Layer}. This layer manages the flow of information from multiple systems, such as IoT devices, Enterprise Resource Planning (ERP) systems, logistics databases, and even publicly available data, and brings them together into a unified graph structure. Given the complexity and multi-layered nature of global supply chains, we ensure that this data is consistently cleaned, standardized, and formatted through robust data processing pipelines and real-time streaming frameworks. By doing so, we enable continuous, real-time monitoring and analysis, setting the stage for more informed and proactive decision-making across the entire supply chain network.

\subsubsection{Overview of Key Data Sources}
Here, we discuss the key data sources that play a vital role in Supply Chain Management:

\begin{enumerate}
\item\textbf{IoT Devices:}
IoT technology has greatly transformed how we collect data in supply chains, providing detailed, real-time insights through advanced devices \citep{phase2018using}. For example, GPS trackers help us track assets and monitor delivery schedules accurately, while RFID and NFC tags automate inventory management, reducing errors in shipment tracking and storage \citep{coskun2011near,domdouzis2007radio}. Smart sensors also play a crucial role in monitoring important environmental conditions such as temperature, humidity, power, and electric current, which are especially important for preserving sensitive items like pharmaceuticals \citep{ramakrishnan_raman__2023}. Additionally, Programmable Logic Controllers (PLCs) provide real-time data that updates Manufacturing Execution Systems (MES) and Digital Twins (DTs), ensuring accurate and up-to-date models of our supply chain processes \citep{ghosh2020novel}. QR codes, which are widely used and cost-effective, also assist in inventory tracking and management. When integrated into Digital Twin frameworks, IoT devices help us detect anomalies in real time and allow us to respond proactively by analyzing the incoming data. Leading companies like Siemens, IBM, Microsoft, PTC, and Tesla have adopted digital twin solutions to predict equipment failures, optimize maintenance, and synchronize production data \citep{aheleroff2021digital,lu2020digital}. However, we must also consider the security and privacy concerns that come with IoT devices. To ensure the safe implementation of these technologies, it is important to implement robust encryption protocols and secure data-sharing frameworks that comply with relevant regulations.

\item\textbf{Enterprise Resource Planning (ERP) Systems:}  
ERP systems bring together key business functions, providing us with valuable data about inventory levels, production schedules, and sales performance. This structured data is essential for creating a clear and detailed view of supply chain dynamics. For instance, ERP systems help us track inventory shortages, align production schedules with real-time demand, and manage sales trends, enabling businesses to respond proactively. Their integration with manufacturing execution systems and state-task networks further enhances their value by optimizing resource allocation \citep{khan2020improving}.

\item\textbf{Logistics Databases:}  
Logistics databases store important information on transportation routes, carrier schedules, and vehicle capacities. These databases help us optimize routes by considering factors like distance, traffic, and delivery windows, ensuring shipments are scheduled in a cost-effective and timely manner. Integrating these databases with GPS systems allows us to track shipments in real time, which improves visibility and customer service. Additionally, historical data in these systems helps us analyze performance and identify trends, such as recurring bottlenecks or delays \citep{c__y__lam__2019}.

\item\textbf{Public Data:}  
External data, such as weather conditions and geopolitical events, plays a crucial role in managing risks proactively. For example, real-time weather data helps us anticipate disruptions due to natural disasters, while geopolitical data highlights potential risks like trade conflicts or political instability \citep{caldara2022measuring}. These external data sources, which are often provided by government agencies, commercial services, or news outlets, complement our internal data, helping us form a more comprehensive understanding of the supply chain.
\end{enumerate}

\subsubsection{Enhanced Data Processing Pipelines}
The Extract, Transform, Load (ETL) process is the core of our Data Integration Layer, enabling us to transform raw data from various sources into a standardized and actionable format \citep{bansal2014towards}. This ensures that the data we feed into our graph-based Digital Twin (DT) is high-quality and ready for analysis.

\noindent \begin{enumerate} 
\item \textbf{Extraction Phase:} In this phase, we gather raw data from multiple systems, such as IoT sensors tracking shipment conditions, ERP systems recording inventory and production metrics, logistics databases with routing and carrier schedules, and external data from public APIs like weather forecasts and geopolitical updates \citep{robert_s__woodley__2011}. We ensure that we can handle various data formats and protocols, ensuring comprehensive collection without losing critical information. 

\item \textbf{Transformation Phase:} Here, we focus on cleaning and transforming the data. We address issues like missing values, duplicates, and inconsistencies across different formats \citep{deepak_r__n__varma__2023}. We apply normalization techniques to bring the data into a consistent format that allows it to be integrated smoothly. For example, we align data from IoT devices monitoring temperature and humidity with inventory records in ERP systems to detect environmental changes that could impact product quality \citep{khan2020improving}. 

\item \textbf{Load Phase:} In the final phase, the cleaned and transformed data is loaded into a centralized repository or data warehouse. This repository is designed to handle the high throughput required for graph-based modeling. It stores the data in a way that makes it easy to retrieve and process quickly during graph construction and analysis. We ensure that the loading process is efficient, so data is available in near real-time, allowing for timely updates \citep{mahmudur_khan_2024}. 
\end{enumerate}

\begin{table}[h]
\centering
\caption{Key Data Sources and Their Role in the ETL Process}
\label{tab:data_sources_etl}
\begin{tabularx}{\textwidth}{|>{\raggedright\arraybackslash}X|>{\raggedright\arraybackslash}X|>{\raggedright\arraybackslash}X|>{\raggedright\arraybackslash}X|}
\hline
\textbf{ETL Phase}       & \textbf{Data Source}       & \textbf{Description}                                            & \textbf{Example Activities}                                  \\ \hline

\multirow{4}{*}{\textbf{Extraction Phase}} 
                         & IoT Devices               & Real-time asset tracking, environmental monitoring.             & GPS tracking, RFID for inventory, sensor data collection.    \\ \cline{2-4}
                         & ERP Systems               & Structured business data on inventory and production.           & Inventory reports, production schedules, sales metrics.      \\ \cline{2-4}
                         & Logistics Databases       & Routing, scheduling, and capacity data.                         & Route optimization, carrier schedules, historical analysis.  \\ \cline{2-4}
                         & Public Data               & External insights: weather, geopolitics, regulations.           & Forecasts, trade rules, political stability data.            \\ \hline

\textbf{Transformation Phase} & All Data Sources          & Cleans, standardizes, aligns data from diverse sources.          & Remove duplicates, fill gaps, normalize formats.             \\ \hline

\textbf{Load Phase}      & Centralized Repository    & Optimized storage for real-time graph-based modeling.           & Load into data warehouses or NoSQL databases.                \\ \hline

\end{tabularx}
\end{table}

We design the ETL process to manage the complexities of large-scale supply chains, ensuring that data from various and geographically spread sources is consistently transformed into a unified framework. By adhering to strict data quality standards, our ETL pipeline ensures the Digital Twin's analysis and predictions are reliable and accurate. This attention to detail is critical for maintaining the effectiveness of the system in providing real-time, actionable insights.

\subsubsection{Advanced Real-Time Streaming Frameworks}
To handle the demands of real-time data ingestion and processing, we utilize advanced streaming frameworks like Apache Kafka. Kafka is known for its scalability, fault tolerance, and low latency, making it ideal for continuous updates in global supply chains \citep{thein2014apache}. It efficiently manages large data streams from IoT devices, ERP systems, and external APIs, processing up to 1,300 events per second with latencies as low as 6–15 milliseconds \citep{k__padmanaban__2024}. This ensures that time-sensitive data, like live shipment tracking and production updates, is ingested seamlessly, reducing the risk of downtime. 

To strengthen Kafka’s capabilities, we integrate it with tools like Apache Spark for data transformation and Cassandra for NoSQL storage, forming a strong data pipeline \citep{jay_oza__2024}. We also complement it with frameworks like Apache Flink and Pulsar. Flink, in particular, is praised for its advanced stream processing and lower latency in high-throughput situations \citep{zhang2021research}. These tools enable real-time cleaning, enrichment, and analysis of incoming data streams, such as aggregating sensor data from transit vehicles to ensure sensitive goods are stored under the right conditions.

Kafka also supports event-driven processing, where deviations from expected metrics—such as inventory shortages or delays—trigger immediate alerts and system actions to keep the Digital Twin up to date \citep{thein2014apache}. Comparative studies show Kafka's resilience under heavy loads, while Flink may perform better in certain high-performance scenarios, giving us flexibility to optimize configurations based on specific needs \citep{m__haseeb_javed__2017}. Together, these tools form the foundation for real-time analytics and operational responsiveness, ensuring we can effectively manage data and respond to changes quickly.

\begin{table}[h]
\centering
\caption{Advanced Frameworks and IoT Contributions in Supply Chain Data Integration}
\label{tab:streaming_iot}
\begin{tabular}{|p{2cm}|p{3.5cm}|p{5cm}|p{5cm}|}
\hline
\textbf{Category} & \textbf{Technology/Device} & \textbf{Functionality} & \textbf{Applications in Supply Chain} \\ \hline
\multirow{4}{*}{IoT Devices} 
    & GPS Trackers   & Real-time location tracking of assets.   & Shipment visibility, optimized routing, and delivery time accuracy. \\ \cline{2-4}
    & RFID/NFC Tags  & Automates inventory updates and reduces errors.  & Warehouse stock tracking, automated inventory counts, and shipment validation. \\ \cline{2-4}
    & Smart Sensors  & Monitors environmental conditions (e.g., temperature, humidity).  & Ensures compliance for sensitive goods like pharmaceuticals and perishables. \\ \cline{2-4}
    & QR Codes       & Low-cost inventory and asset tracking solution.  & Facilitates quick inventory checks and product identification. \\ \hline
\multirow{4}{*}{Streaming} 
    & Apache Kafka   & High-throughput real-time data ingestion with fault tolerance.  & Processes live sensor feeds, shipment tracking, and external updates. \\ \cline{2-4}
    & Apache Flink   & Low-latency stream processing for event-driven applications.  & Detects anomalies (e.g., delays or environmental deviations) in transit data. \\ \cline{2-4}
    & Apache Spark   & Batch and streaming data transformation with scalable architecture.  & Aggregates IoT data with ERP records for centralized analysis. \\ \cline{2-4}
    & Cassandra      & NoSQL database optimized for fast, distributed storage of large datasets.  & Stores historical IoT data for long-term performance analysis and trend insights. \\ \hline
\end{tabular}
\end{table}

\subsubsection{Challenges in the Data Integration Layer}
In the Data Integration Layer, which plays a crucial role in the functionality of graph-based Digital Twins, we encounter several challenges that can impact its effectiveness. One of the main issues is dealing with missing or incomplete data. This is common in supply chains where information can be fragmented or inconsistent across various sources. For instance, inventory levels reported by ERP systems may not always match real-time sensor data from warehouses, which requires us to use advanced imputation techniques to fill these gaps and ensure the data is complete \citep{rui_wu__2022}. 

Another challenge we face is ensuring interoperability. Since global supply chains rely on diverse data formats, communication protocols, and system architectures, integrating systems like GPS trackers and ERP platforms can be complex. We can address this issue by using middleware solutions and standardized data schemas, which help different systems communicate more effectively \citep{wang2019method}. 
Data security and privacy are also significant concerns. Sensitive information, such as shipment schedules or supplier contracts, is vulnerable to breaches and unauthorized access during transmission and storage. To mitigate these risks, we must implement robust encryption methods and secure access protocols to protect data integrity and ensure compliance with regulations like GDPR \citep{pantlin2018supply}. 
Scalability presents an additional challenge, especially as supply chains expand and the volume of data from IoT devices, logistics platforms, and external sources increases. We need scalable infrastructures, such as cloud-based architectures, to handle this growing complexity without compromising performance or incurring excessive costs \citep{favour_amarachi_ezeugwa_2024}.

\subsection{Graph Construction Module Function}
\begin{figure}
    \centering
    \includegraphics[width=\linewidth]{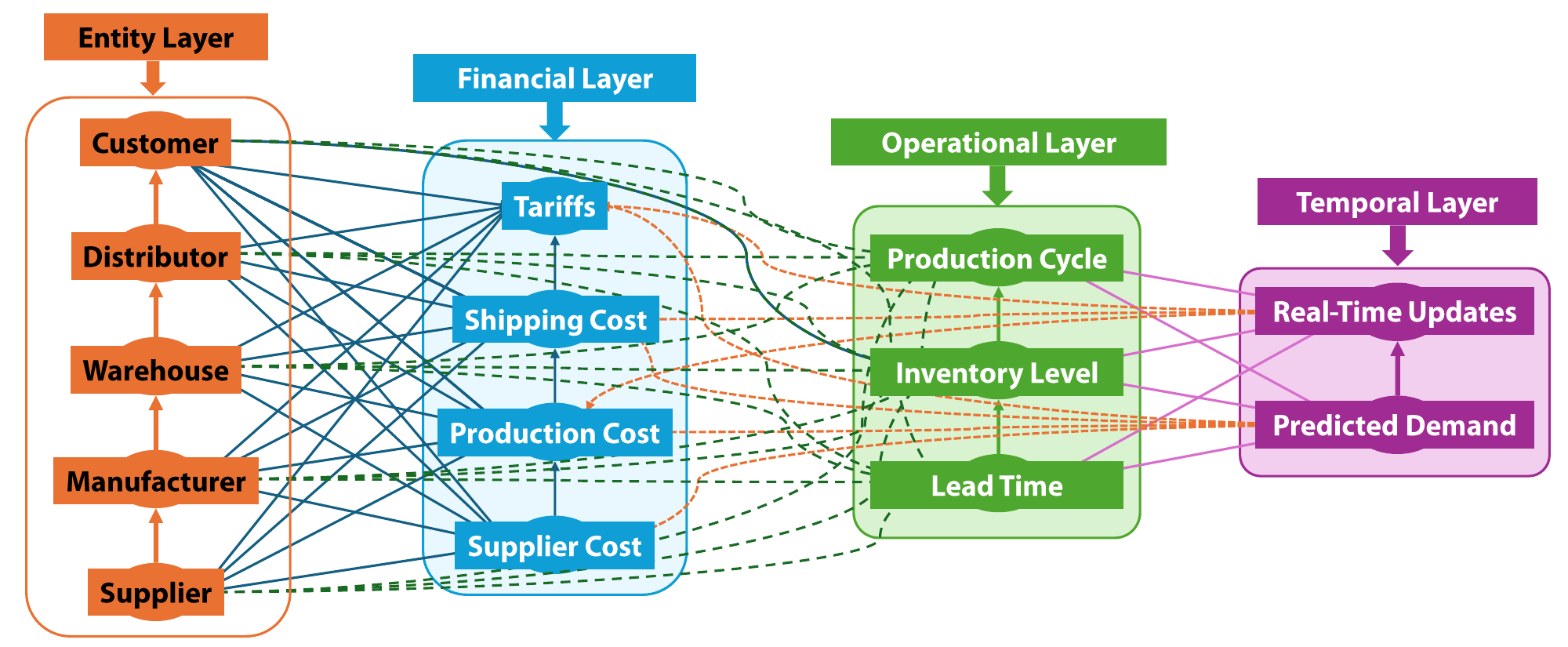}
    \caption{Multilayer Graphs in Supply Chain Optimization}
    \label{fig:5_MultiLayerSCG}
\end{figure}

Graph Construction Module plays a critical role in transforming the integrated and preprocessed data from the Data Integration Layer into a graph representation that effectively models the supply chain network, as demostrated in Figure \ref{fig:5_MultiLayerSCG}. This module forms the foundation for advanced analysis and decision-making by organizing supply chain entities and their interconnections in a graph format. Building on the standardized data we obtained in the first layer, we translate the complex, multi-tiered relationships within the supply chain into a dynamic, scalable graph structure. In doing so, we combine domain knowledge with computational techniques to capture the intricate dependencies and interactions among suppliers, manufacturers, distributors, and customers. This process ensures that we have a comprehensive representation of the supply chain that supports informed decision-making and enhances operational efficiency.

\subsubsection{Nodes: Representing Supply Chain Entities}
In our graph, nodes represent key supply chain entities, each capturing the distinct roles and functions of various stakeholders within the network \citep{wu2023industry}. Suppliers provide essential raw materials or components for production, such as semiconductor chip suppliers in the electronics industry, whose performance directly impacts downstream manufacturing processes. Manufacturers are responsible for converting raw materials into finished products, shaping production schedules, product quality, and cost efficiency through their relationships with suppliers and distributors. Warehouses act as intermediate storage points, where factors like inventory levels, throughput, and handling times become critical for overall efficiency. Distributors connect warehouses or manufacturers to retailers or end customers, ensuring timely and efficient product delivery. Customers, whether individuals or businesses, are modeled as terminal nodes, offering valuable insights into demand patterns, order fulfillment times, and overall satisfaction. Each node defines a role within the network and contributes key operational metrics—like production capacity, inventory levels, or demand volume—which are essential for informed decision-making \citep{prem_prakash_mishra__2017}. 

\begin{tcolorbox}[colback=magenta!5!white,colframe=black,boxsep=0pt,top=4pt,bottom=4pt,left=3pt,right=3pt]
Mathematically, we can define the set of nodes as \(N = \{n_1, n_2, \dots, n_k\}\), where \(n_i\) represents a specific supply chain entity such as a supplier, manufacturer, warehouse, distributor, or customer. The node attributes could be represented as \(A(n_i) = \{a_1, a_2, \dots, a_m\}\), capturing operational metrics relevant to each entity, like capacity, inventory levels, and demand.
\end{tcolorbox}

\subsubsection{Edges: Representing Relationships Between Entities}
In our graph, edges represent the relationships and interactions between supply chain entities, capturing the flow of goods, information, or finances throughout the network \citep{victoralex_2022}. These edges are fundamental to the network’s structure, reflecting transportation routes, supplier contracts, and information exchanges. Transportation routes connect suppliers, warehouses, and distributors through physical logistics pathways like road, rail, air, or sea, with key attributes such as distance, travel time, and cost. Supplier contracts model procurement agreements, highlighting delivery schedules, pricing, and volume commitments between suppliers and manufacturers. Information flows track the exchange of data, such as inventory updates, production plans, and customer orders, enabling synchronization across the network and supporting predictive analysis. These edges are critical for modeling dependencies within the network, especially in cases where disruptions in transportation or delays in information exchange can cause cascading effects \citep{peng_ji__2024}. By applying centrality measures to edges, we can identify key paths whose failure would severely disrupt the supply chain.

\begin{tcolorbox}[colback=magenta!5!white,colframe=black,boxsep=0pt,top=4pt,bottom=4pt,left=3pt,right=3pt]
Mathematically, we define the set of edges as \(E = \{e_1, e_2, \dots, e_m\}\), where each edge \(e_i\) represents a relationship between two nodes, such as a transportation route, contract, or information flow. The edge attributes can be represented as \(A(e_i) = \{a_1, a_2, \dots, a_n\}\), which may include distance, cost, and flow capacity, depending on the nature of the relationship being modeled.
\end{tcolorbox}

\begin{table}[h]
\centering
\caption{Nodes, Edges, and Graph Attributes in Supply Chain Representation}
\label{tab:nodes_edges_attributes}
\begin{tabular}{|p{2cm}|p{3cm}|p{6cm}|p{4.5cm}|}
\hline
\textbf{Type}       & \textbf{Sub-Type}            & \textbf{Description}                                                           & \textbf{Attributes}                                                                                                 \\ \hline

\multirow{5}{*}{\textbf{Nodes}} 
                    & Suppliers                   & Provide raw materials or components for production.                            & Availability, pricing, delivery timelines.                                                                         \\ \cline{2-4}
                    & Manufacturers               & Transform raw materials into finished goods.                                   & Production capacity, lead times, defect rates.                                                                     \\ \cline{2-4}
                    & Warehouses                  & Serve as intermediary storage points.                                          & Storage capacity, inventory levels, handling time.                                                                 \\ \cline{2-4}
                    & Distributors                & Ensure goods reach retailers or customers efficiently.                         & Delivery schedules, geographic coverage, vehicle capacity.                                                         \\ \cline{2-4}
                    & Customers                   & Represent end-users or businesses purchasing finished products.                & Demand volume, order fulfillment rates, satisfaction scores.                                                       \\ \hline

\multirow{4}{*}{\textbf{Edges}} 
                    & Transportation Routes       & Connect suppliers, warehouses, and distributors for physical logistics.        & Distance, travel time, fuel costs, tariffs.                                                                        \\ \cline{2-4}
                    & Supplier Contracts          & Model procurement agreements between suppliers and manufacturers.              & Delivery schedules, pricing, volume commitments.                                                                   \\ \cline{2-4}
                    & Information Flows           & Represent data exchanges across the network.                                   & Inventory updates, production plans, customer orders.                                                              \\ \cline{2-4}
                    & Financial Flows             & Track monetary transactions and payment schedules.                             & Transaction value, payment delays, penalties.                                                                      \\ \hline

\multirow{4}{*}{\textbf{Attributes}} 
                    & Cost                        & Represents financial metrics across nodes and edges.                           & Production costs, shipping fees, penalties for delays.                                                             \\ \cline{2-4}
                    & Time                        & Captures durations related to supply chain processes.                          & Lead times, shipping durations, production cycles.                                                                 \\ \cline{2-4}
                    & Reliability                 & Measures consistency and dependability of nodes and edges.                     & On-time delivery rates, equipment failure rates.                                                                   \\ \cline{2-4}
                    & Capacity                    & Indicates throughput limits for nodes and edges.                               & Warehouse storage, vehicle cargo limits.                                                                           \\ \hline

\end{tabular}
\end{table}

\subsubsection{Attributes: Enhancing Graph Representation}
In our graph representation, we enhance both nodes and edges with attributes that provide deeper operational insights and enable more advanced analyses. Cost attributes capture key financial metrics, such as production costs, shipping expenses, and penalties for delays. For example, transportation edges may include fuel costs and tariffs, while supplier nodes account for the price of raw materials. Time attributes represent durations like shipping times, lead times, and production cycle lengths, which are especially important for time-sensitive goods like pharmaceuticals \citep{rhea_m_thomas__2024}. Reliability attributes help us quantify the consistency and dependability of nodes and edges, using metrics such as on-time delivery rates, equipment failure rates, and historical performance data. Capacity attributes define throughput limits for nodes, such as the storage capacity of warehouses, and edges, like the cargo limits of vehicles, ensuring the supply chain can meet demand without facing bottlenecks or inefficiencies. By combining these various attributes, we create a more robust and comprehensive graph that provides a clear and actionable representation of the entire supply chain network.

\begin{tcolorbox}[colback=magenta!5!white,colframe=black,boxsep=0pt,top=4pt,bottom=4pt,left=3pt,right=3pt]
Mathematically, we represent the attributes of nodes and edges as follows: For each node \(n_i\), the attributes can be expressed as \(A(n_i) = \{a_1, a_2, \dots, a_p\}\), where each \(a_j\) corresponds to a specific attribute like cost, reliability, or capacity. Similarly, for each edge \(e_i\), the attributes are represented as \(A(e_i) = \{b_1, b_2, \dots, b_q\}\), where each \(b_k\) represents an edge attribute such as cost, time, or capacity. These sets of attributes help us capture the complex and dynamic nature of the supply chain network.
\end{tcolorbox}

In our Graph Construction Module, we use different types of graphs—static, dynamic, and multi-layer—to model supply chain networks. Each type offers unique insights into various aspects of the supply chain, providing us with complementary perspectives. Static graphs help us understand the fixed relationships between entities, while dynamic graphs capture the evolving nature of the supply chain over time, accounting for changes like inventory levels or demand fluctuations. Multi-layer graphs allow us to represent multiple types of interactions within the network, such as transportation, procurement, and information flow, all in one unified structure. By using these different graph types together, we gain a more comprehensive understanding of the network’s structure, dynamics, and interdependencies, which significantly enhances our ability to analyze and make decisions in the context of graph-based Digital Twins (DTs).

\begin{table}[h]
\centering
\caption{Simplified Graph Types, Characteristics, Applications, and Use Case Examples}
\label{tab:simplified_graph_types}
\begin{tabular}{|p{1.5cm}|p{2.5cm}|p{3.5cm}|p{3.5cm}|p{4cm}|}
\hline
\textbf{Graph Type}       & \textbf{Purpose}                                 & \textbf{Key Features}                            & \textbf{Applications}                       & \textbf{Use Case Examples}                                   \\ \hline

\textbf{Static Graphs}    & Snapshot of supply chain at one time.            & Topology, bottlenecks, critical nodes.           & Inventory, route optimization.              & Identify critical suppliers disrupting the network.          \\ \hline

\textbf{Dynamic Graphs}   & Captures time-based changes.                     & Tracks demand, disruptions, cascades.            & Forecast trends, adjust in real time.       & Simulate factory shutdown impact on inventory/delivery.      \\ \hline

\textbf{Multi-Layer Graphs} & Shows multiple relationships in layers.         & Combines material, info, financial flows.        & Optimize dependencies, improve collaboration. & Analyze payment delays impacting material shipments.         \\ \hline

\end{tabular}
\end{table}

\subsubsection{Static Graphs: Capturing Fixed Relationships}
Static graphs provide us with a snapshot of the supply chain network at a specific point in time, offering a clear view of its structure and the relationships between different entities \citep{wasi}. In these graphs, nodes represent supply chain entities like suppliers, manufacturers, warehouses, distributors, and customers, while edges capture relationships such as material flows, transportation links, and contracts. Static graphs are particularly useful for analyzing the network's structure, helping us identify key nodes, bottlenecks, and critical paths. For example, a static graph could show how raw materials flow from suppliers to manufacturers and then to retailers, revealing important nodes that influence the network's efficiency \citep{junhong_cui__2024}. This helps us optimize inventory management, spot potential risks, and plan better transportation routes. Static graphs serve as a starting point, providing us with essential structural insights before we introduce time-based or layered complexities. 

\begin{tcolorbox}[colback=magenta!5!white,colframe=black,boxsep=0pt,top=4pt,bottom=4pt,left=3pt,right=3pt]
Mathematically, a static graph can be represented as \( G = (V, E) \), where \( V \) is the set of nodes (supply chain entities) and \( E \) is the set of edges (relationships between entities).
\end{tcolorbox}

\subsubsection{Dynamic (Time-Evolving) Graphs: Reflecting Changes Over Time}
Dynamic graphs take the concept of static graphs further by incorporating changes over time, offering a more realistic view of the supply chain’s evolution \citep{ke_cheng__2024}. In dynamic graphs, both nodes and edges can change, reflecting shifts like demand fluctuations, production schedules, and disruptions such as natural disasters or supplier failures. This time-evolving perspective allows us to track real-time changes, forecast future trends, and build resilience in the supply chain. For example, a dynamic graph could show how inventory levels fluctuate across warehouses, identifying patterns like seasonal stock increases or shortages \citep{joo_santos__2024}. It can also simulate the impact of disruptions, such as a route closure, and help us proactively reroute shipments. Dynamic graphs are particularly useful for predicting demand and assessing the consequences of disruptions, offering real-time insights that enable quick adjustments to changing conditions \citep{sarimveis2008dynamic}. By analyzing these temporal patterns and past data, dynamic graphs help with predictive decision-making. 

\begin{tcolorbox}[colback=magenta!5!white,colframe=black,boxsep=0pt,top=4pt,bottom=4pt,left=3pt,right=3pt]
Mathematically, a dynamic graph can be represented as \( G(t) = (V(t), E(t)) \), where \( V(t) \) and \( E(t) \) are the sets of nodes and edges at time \( t \), and the graph evolves over time, capturing the supply chain’s changing state. Each node and edge may have a time-dependent attribute, such as inventory levels or transportation costs, represented as \( \text{attr}_{v}(t) \) for nodes and \( \text{attr}_{e}(t) \) for edges.
\end{tcolorbox}

\subsubsection{Multi-Layer Graphs}
Multi-layer graphs allow us to model the supply chain through several interconnected layers, each highlighting a different type of relationship or interaction. These layers include material flow, which tracks the movement of raw materials, components, and finished goods; information flow, which covers data exchanges like inventory updates, shipment notifications, and production schedules; financial flow, which details monetary transactions and payment schedules; and social interactions, which map the relationships and collaborations among stakeholders to promote communication and cooperation within the network \citep{a2018unfolding}. By combining these layers, multi-layer graphs offer a comprehensive view of the supply chain, helping us uncover hidden dependencies, such as how supplier payment delays can affect production timelines. They also provide insights into social interactions, helping us identify key influencers or communication bottlenecks that could improve collaboration. These graphs can further support advanced optimization methods, like using Graph Neural Networks (GNNs) to explore interdependencies and find ways to enhance efficiency, providing a more integrated and complete perspective for managing the supply chain \citep{edward_elson_kosasih__2021}. 

\begin{tcolorbox}[colback=magenta!5!white,colframe=black,boxsep=0pt,top=4pt,bottom=4pt,left=3pt,right=3pt]
Mathematically, a multi-layer graph can be represented as \( G = (V, E) \), where \( V \) is the set of nodes and \( E \) represents the set of edges across multiple layers. Each layer \( L_i \) corresponds to a different type of flow or relationship, such as material flow, information flow, or financial flow, and can be described as \( G_{L_i} = (V, E_{L_i}) \), where \( E_{L_i} \) denotes the edges specific to layer \( L_i \). The layers are interconnected through shared nodes, allowing us to capture the complex interdependencies between different types of flows.
\end{tcolorbox}

The main goal of our Graph Construction Module is to transform the integrated and preprocessed data from the Data Integration Layer into a detailed, weighted graph that captures the complex dynamics of the supply chain network. This module is designed to overcome the limitations of traditional, linear, and siloed models by offering a multidimensional, interconnected view of supply chain entities and their relationships. In our graph, nodes represent key entities like suppliers, manufacturers, distributors, warehouses, and customers, each enhanced with attributes such as inventory levels, production capacities, and demand requirements. These nodes are connected by directed edges that model essential relationships like material flow, financial transactions, and information exchange, creating a dynamic structure that reflects real-time operations \citep{prem_prakash_mishra__2017}. We also assign weights to the edges—such as cost, lead time, quantity, and risk—which help quantify the factors influencing supply chain performance. For instance, material flow edges could be weighted based on transportation costs or shipping times, while financial flow edges might represent transaction amounts or payment delays, offering a detailed view of the operational interdependencies \citep{victoralex_2022}. By incorporating live data from ERP systems, logistics databases, and external sources, we ensure that these weights adapt to real-time changes, such as weather disruptions or geopolitical events. Unlike traditional methods, our weighted graph provides deeper insights, allowing managers to simulate different scenarios, identify bottlenecks, and evaluate strategies with precision. Using advanced graph algorithms, such as centrality measures to identify crucial nodes or shortest-path calculations to optimize routes, this approach turns static data into a dynamic decision-support system. Ultimately, the weighted graph becomes a real-time tool that supports proactive decision-making, boosts supply chain resilience, and reveals optimization opportunities, laying the foundation for future analytics and enhancing supply chain management.

\subsection{Simulation and Analysis Engine}
\begin{figure}
    \centering
    \includegraphics[width=\linewidth]{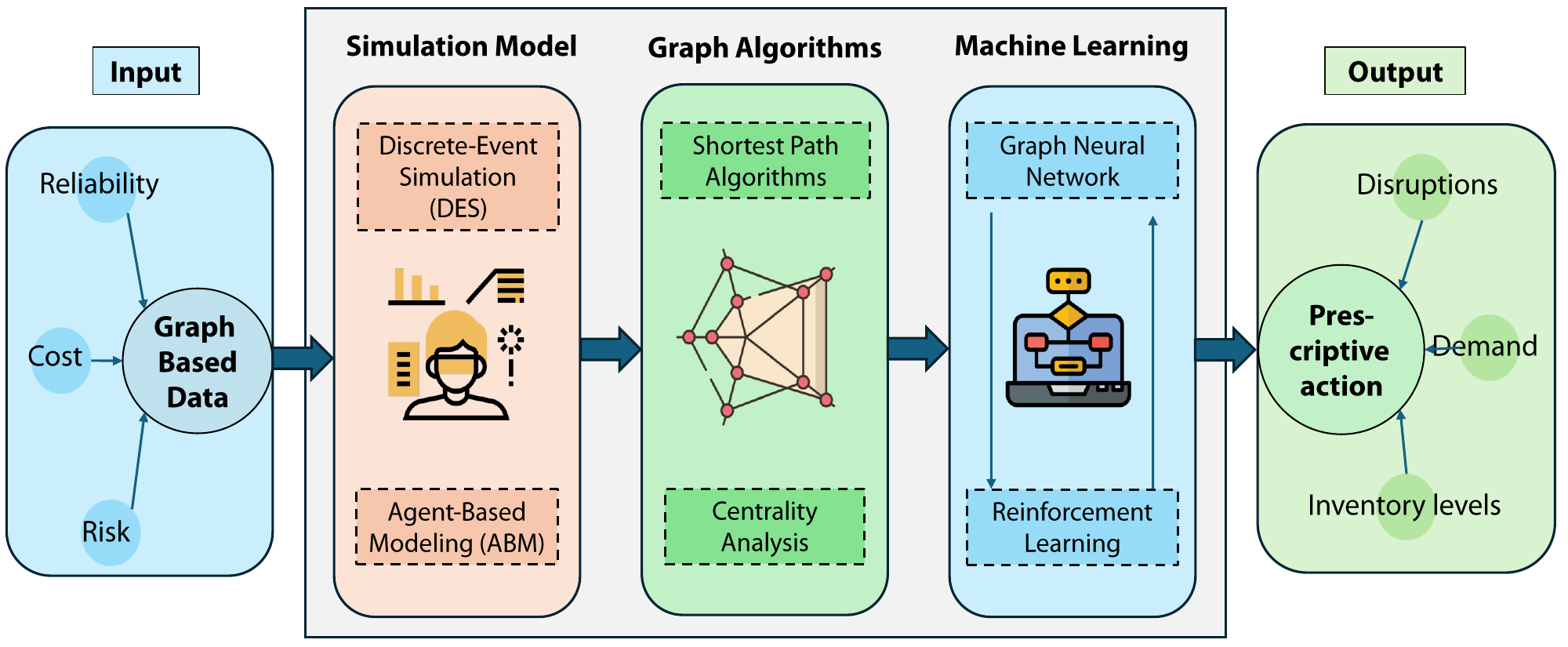}
    \caption{Workflow and Components of Simulation and Analysis Engine}
    \label{fig:6_Simulation_and_Analysis_Engine}
\end{figure}

Simulation and Analysis Engine, as the third layer in our architecture, serves as the strategic heart of the graph-based Digital Twin (DT) framework. This module brings together advanced computational techniques to simulate complex scenarios, predict potential outcomes, and analyze vulnerabilities within the supply chain. Building on the graph-structured data provided by the earlier layers, it transforms static supply chain models into dynamic, interactive systems that offer real-time insights and decision support, as illustrated in Figure \ref{fig:6_Simulation_and_Analysis_Engine}. By utilizing state-of-the-art simulation models, predictive analytics, and scenario-based evaluations, this layer gives organizations the tools to anticipate disruptions, optimize their operations, and improve overall network resilience. This is especially crucial in today’s globalized supply chains, where uncertainty, interdependencies, and unexpected challenges require proactive and flexible solutions. Through its dynamic approach, the Simulation and Analysis Engine helps turn supply chains from reactive systems into resilient, agile networks that can navigate complexities effectively and with precision.

\subsubsection{Simulation Models}
Our \textbf{Simulation and Analysis Engine} relies on a diverse set of simulation models, each tailored to capture different aspects of supply chain dynamics. One of the fundamental approaches we use is Discrete-Event Simulation (DES), which models supply chain operations as a series of distinct events occurring over time. This method helps us analyze disruptions, such as a sudden halt in a supplier’s production, by showing how it affects inventory levels, customer service rates, and overall material flow \citep{maroua_soumatia__2024}. For example, DES can quantify how long a supply chain can sustain customer demands before stock shortages occur due to a delay. Such insights allow managers to evaluate risk profiles in real-time and develop adaptive strategies to minimize the ripple effects of these disruptions \citep{mark_holmes_2024}. By providing a granular view of bottlenecks and supply shortages, DES plays a key role in ensuring that supply chains remain efficient and responsive to uncertainties.

Beyond DES, we incorporate Agent-Based Modeling (ABM) and System Dynamics Modeling, which take a more systemic approach to analyzing supply chains. These models represent different supply chain entities—such as manufacturers, logistics providers, and retailers—as independent decision-makers interacting within a complex network. ABM is particularly useful in simulating real-world scenarios like how logistics providers react to natural disasters, while system dynamics modeling helps us understand long-term consequences, such as shifts in market demand due to a supply chain disruption \citep{liming_xu__2023}. By running “what-if” simulations, we can test different strategies, such as rerouting shipments or adjusting production schedules, to evaluate their effectiveness in preventing delays and improving resilience \citep{silja_meyernieberg__2014}. These insights enable businesses to proactively reconfigure their supply chain operations rather than reacting to disruptions after they occur.

To further enhance our ability to assess risks, we incorporate stress testing methods that identify vulnerabilities within the supply chain network. For instance, entropy-based assessments help us detect critical nodes—such as major distribution hubs—that, if disrupted, could severely impact overall operations \citep{ziqiang_zeng__2024}. We also integrate hybrid simulation techniques, such as combining Susceptible-Infected-Recovered (SIR) models with agent-based simulations. This approach proved particularly valuable during the COVID-19 pandemic, where it demonstrated how early interventions and diversified supplier partnerships could effectively mitigate cascading risks \citep{shubham_rajendra_ekatpure_2024}. Additionally, optimization techniques like evolutionary computation allow us to model worst-case scenarios and develop robust countermeasures. By balancing trade-offs between cost, time, and resilience, these simulations ensure that our mitigation strategies are both practical and aligned with business goals. Through this comprehensive and predictive approach, we transform supply chain management from a reactive process into a proactive, strategically driven discipline, helping businesses navigate uncertainties with agility and precision.

\subsubsection{Graph algorithms}
In our Simulation and Analysis Engine, we leverage advanced graph algorithms to improve supply chain efficiency and resilience. These algorithms operate on the weighted graph built in the previous layer, allowing us to optimize routes, uncover interdependencies, and identify critical components within the network. By applying methods like shortest path analysis, community detection, and centrality measures, we gain actionable insights that help mitigate risks and enhance strategic decision-making \citep{jan2019analysis}. Shortest path algorithms, such as Dijkstra’s algorithm, are particularly useful for finding the most efficient transportation routes based on factors like distance, cost, or lead time \citep{makariye2017towards}. For example, a distributor can use Dijkstra’s algorithm to determine the most cost-effective delivery route from a warehouse to a construction site \citep{ittrotul_muyammina__2024}. When dealing with graphs that include negative weights—such as scenarios involving penalties or rebates—the Bellman-Ford algorithm provides an effective solution, while the Floyd-Warshall algorithm helps compute shortest paths for all node pairs in dense networks, improving overall logistics efficiency \citep{michael_j__bannister__2012}.

Beyond route optimization, community detection algorithms help us uncover clusters of interconnected supply chain entities, such as groups of suppliers, manufacturers, and distributors that collaborate within a specific region. Understanding these clusters is crucial because disruptions in one part of the community—such as a supplier shutting down—can cause cascading effects throughout the network. By applying modularity optimization techniques, we can detect these interdependencies and identify structural weaknesses, allowing us to build more resilient supply chain networks \citep{yingqiu_zhu__2024}. Additionally, centrality analysis helps us pinpoint the most critical nodes and links within the supply chain. Metrics like degree, betweenness, and closeness centrality allow us to rank the influence of various components. For example, a distribution center with high betweenness centrality might handle a significant volume of goods, making it a key point for network efficiency, while a transportation link with high edge centrality might be essential for connecting major suppliers and manufacturers \citep{i__shvarts__2022}. By identifying these crucial elements, we can implement targeted interventions, such as securing vital distribution centers or fortifying key transportation routes, to enhance overall robustness.

Together, these graph algorithms transform our Digital Twin into a powerful tool for predictive and prescriptive analytics. By optimizing transportation flows, anticipating disruptions, and identifying vulnerabilities, we enable supply chains to shift from reactive management to a proactive, strategic approach. This level of adaptability ensures that businesses remain resilient in the face of disruptions, helping them navigate the challenges of today’s complex and volatile global markets.

\subsubsection{Machine Learning Models}
Machine learning, particularly Graph Neural Networks (GNNs), is transforming supply chain management by utilizing the naturally structured nature of supply chain data \citep{ahn2024gnn}. These models help us predict demand patterns, manage disruptions, and optimize network flows by capturing the complex relationships between supply chain components—such as suppliers, manufacturers, and transportation routes. Since GNNs are specifically designed for graph-structured data, they offer superior decision-making capabilities in dynamic environments. For instance, GNNs have been highly effective in origin-destination (OD) demand prediction for urban traffic systems, integrating hybrid neural network architectures to deliver real-time, disruption-resistant forecasts \citep{haodong_ma__2023}. Similarly, in power distribution networks, GNNs have been employed for optimal power flow (OPF) prediction, reducing computational costs while improving system efficiency. When applied to supply chains, these techniques enable us to optimize transportation routes, reduce operational costs, and minimize lead times, providing organizations with a more agile and data-driven approach to logistics management \citep{thuan_phamh__2024}.

Beyond GNNs, several other machine learning models contribute to supply chain optimization. Convolutional Neural Networks (CNNs)—though typically used for image analysis—can be adapted to identify spatial patterns in warehouse layouts and transportation routes, improving logistics efficiency \citep{yin2022retracted}. Recurrent Neural Networks (RNNs), well-known for handling time-series data, are valuable for forecasting inventory fluctuations, predicting demand surges, and optimizing replenishment schedules based on past trends \citep{kiuchi2024recurrent}. Support Vector Machines (SVMs) excel in classification tasks, allowing us to analyze supplier reliability and categorize potential risks in logistics networks \citep{zhang2021firefly}. Additionally, ensemble learning techniques, such as Random Forests and Gradient Boosting Machines (GBM), combine multiple models to enhance the accuracy of supplier performance evaluations and customer demand predictions \citep{theoretical_evaluation_of_ensemble_machine_learning_techniques_2023}. Together, these machine learning models form a powerful toolkit that enables us to improve supply chain resilience, enhance operational efficiency, and make data-driven decisions in complex, fast-changing environments.

Reinforcement Learning (RL) offers a powerful, adaptive approach to decision-making in dynamic supply chain environments, learning optimal strategies through trial and error while continuously adjusting to real-time data \citep{franccois2018introduction}. This makes RL particularly effective for inventory management, transportation optimization, and resource allocation in interconnected networks. Key RL techniques, such as Deep Q-Networks (DQNs) and Proximal Policy Optimization (PPO), provide distinct advantages. DQNs handle structured decision-making, optimizing inventory levels and transportation routes, while PPO excels in continuous-action environments, dynamically adjusting shipment schedules and resource distribution \citep{oroojlooyjadid2022deep, seungyul_han__2017}. These problems are often modeled as Markov Decision Processes (MDPs), allowing RL to balance short-term and long-term outcomes \citep{marnix_suilen__2024}.
RL also enhances network optimization and risk management. Multi-Agent Reinforcement Learning (MARL) improves routing efficiency by dynamically adjusting paths to reduce congestion and delays \citep{silva2019reinforcement}. Bayesian RL models incorporate uncertainty, ensuring robust decision-making in unpredictable scenarios \citep{jiayu_chen__2024}. Additionally, entropy-based models help stress-test networks, identifying vulnerabilities and recommending resilience strategies. Innovations like Bayesian Markov Decision Processes enhance adaptability in uncertain conditions, while Fourier Controller Networks improve efficiency by recognizing cyclical patterns. Together, these RL techniques strengthen supply chain flexibility and precision, enabling systems to adapt proactively to evolving challenges.

\begin{table}[ht]
\centering
\caption{Simplified Digital Twin Aspects, Capabilities, and Benefits in the Simulation and Analysis Engine}
\label{tab:simplified_dt_aspects}
\begin{tabular}{|p{3.5cm}|p{4cm}|p{3.5cm}|p{4cm}|}
\hline
\textbf{Digital Twin Aspect}    & \textbf{Digital Twin Capability}                      & \textbf{Enabled Benefits}                    & \textbf{Real-World Outcomes}                                  \\ \hline

\textbf{Simulation Models}      & Simulate disruptions and demand spikes.              & Proactive planning and risk management.      & Rerouting shipments during port closures.                     \\ \hline

\textbf{Graph Algorithms}       & Optimize routes and identify critical nodes.         & Real-time supply chain optimization.         & Reduce lead times and minimize costs.                         \\ \hline

\textbf{Machine Learning Models} & Predict demand and manage disruptions.               & Enhanced resilience and adaptability.        & Detect supplier risks early, avoid downtime.                  \\ \hline

\textbf{Reinforcement Learning}  & Dynamic decision-making in real time.                & Efficient inventory and resource allocation. & Lower storage costs and reduce waste.                         \\ \hline

\textbf{Hybrid Approaches}       & Combine models for comprehensive analysis.           & Improved network robustness.                 & Strengthen critical hubs to prevent cascading failures.        \\ \hline

\end{tabular}
\end{table}

By combining machine learning models like GNNs with RL frameworks, we can move beyond simple predictions to proactive and adaptive supply chain optimization. GNNs help us identify key nodes and connections in a supply chain, while RL systems use this insight to dynamically adjust operations, such as rerouting shipments during disruptions or reallocating resources in real time. This integration creates a resilient and intelligent supply chain ecosystem, equipping businesses to handle uncertainties and improve efficiency in a rapidly changing global market. By leveraging these advanced technologies, we can enhance decision-making, reduce risks, and drive long-term operational success.

\begin{table}[ht]
\centering
\caption{Digital Twin Techniques for Supply Chain Optimization}
\label{tab:combined_dt_techniques}
\scalebox{0.65}{
\begin{tabular}{|p{5cm}|p{4cm}|p{2.5cm}|p{6cm}|p{6cm}|}
\hline
\textbf{Category}               & \textbf{Technique/Model}           & \textbf{Functionality}                              & \textbf{Applications}                               & \textbf{Use Case}                                      \\ \hline

\multirow{3}{*}{\textbf{Graph Algorithms}} 
                                & Shortest Path (Dijkstra)           & Finds efficient routes.                            & Transport and cost optimization.                   & Fastest delivery routes.                               \\ \cline{2-5}
                                & Community Detection                & Identifies clusters of nodes.                     & Managing supplier dependencies.                    & Detecting regional supplier groups.                   \\ \cline{2-5}
                                & Centrality Analysis                & Ranks key nodes or edges.                         & Targeting critical hubs.                           & Securing major distribution centers.                  \\ \hline

\multirow{4}{*}{\textbf{Simulation Models}} 
                                & Discrete Event Simulation          & Models cascading effects.                         & Disruption impact assessment.                      & Evaluating supplier halts.                             \\ \cline{2-5}
                                & Agent-Based Modeling               & Simulates dynamic agents.                         & Testing logistics responses.                       & Adapting to weather disruptions.                      \\ \cline{2-5}
                                & System Dynamics                    & Models feedback over time.                        & Understanding demand trends.                       & Analyzing delays' market impacts.                     \\ \cline{2-5}
                                & Hybrid Models                      & Combines simulation approaches.                   & Managing cascading risks.                          & Pandemic impact studies.                               \\ \hline

\multirow{4}{*}{\textbf{Machine Learning Models}} 
                                & Graph Neural Networks (GNNs)       & Predicts demand, optimizes flows.                & Demand forecasting and flow optimization.          & Managing demand surges.                                \\ \cline{2-5}
                                & Recurrent Neural Networks          & Processes sequential data.                        & Inventory and demand forecasting.                  & Predicting replenishment schedules.                   \\ \cline{2-5}
                                & Support Vector Machines (SVMs)     & Classifies structured data.                       & Supplier risk and reliability analysis.            & Ranking supplier reliability.                          \\ \cline{2-5}
                                & Ensemble Models                    & Combines multiple models.                         & Improving forecasts and trends.                    & Customer purchasing trends.                            \\ \hline

\multirow{6}{*}{\textbf{Reinforcement Learning}} 
                                & Deep Q-Networks (DQNs)             & Discrete action decisions.                       & Inventory, route optimization.                     & Choosing optimal routes.                               \\ \cline{2-5}
                                & Proximal Policy Optimization       & Continuous actions, dynamic adjustments.         & Resource allocation during disruptions.            & Adjusting shipment schedules.                          \\ \cline{2-5}
                                & Multi-Agent RL (MARL)              & Coordinates multiple decision agents.            & Routing and congestion management.                 & Dynamic delivery path adjustments.                     \\ \cline{2-5}
                                & Bayesian RL                        & Adapts to uncertain environments.                & Decision-making under uncertainty.                 & Handling geopolitical tensions.                        \\ \cline{2-5}
                                & Entropy-Based RL                   & Evaluates systemic vulnerabilities.              & Stress-testing supply chains.                      & Identifying critical nodes in the network.             \\ \cline{2-5}
                                & Fourier Controller Networks        & Analyzes cyclic patterns.                        & Resource efficiency in demand cycles.              & Adjusting to seasonal product demand.                  \\ \hline

\end{tabular}}
\end{table}

The first three layers of our graph-based Digital Twin (DT) architecture lay a strong foundation for transforming supply chain management. By integrating real-time data, structured graph models, and advanced analytics, we move from disconnected information to a dynamic, adaptive system. At this stage, our DT is no longer just a static representation—it becomes a living graph that continuously updates with key attributes like cost, time, capacity, and reliability.

With machine learning tools such as GNNs and reinforcement learning, our DT enables both predictive and prescriptive decision-making. For example, when disruptions like supplier failures or port delays occur, the system can dynamically reroute shipments, adjust inventory, and suggest contingency plans \citep{ashraf2024disruption}. These capabilities help us identify bottlenecks, pinpoint critical nodes, and enhance resilience, allowing organizations to move from reactive problem-solving to proactive optimization.
At this point, our DT offers real-time insights and automated actions to improve supply chain efficiency. However, the next and final layer will fully integrate these strategies into real-world operations, making the DT an interactive and agile system. This final step ensures that organizations can seamlessly execute optimized decisions, strengthen resilience, and maintain a competitive edge in an unpredictable global market.

\subsection{Visualization Interface Function}
\begin{figure}
    \centering
    \includegraphics[width=\linewidth]{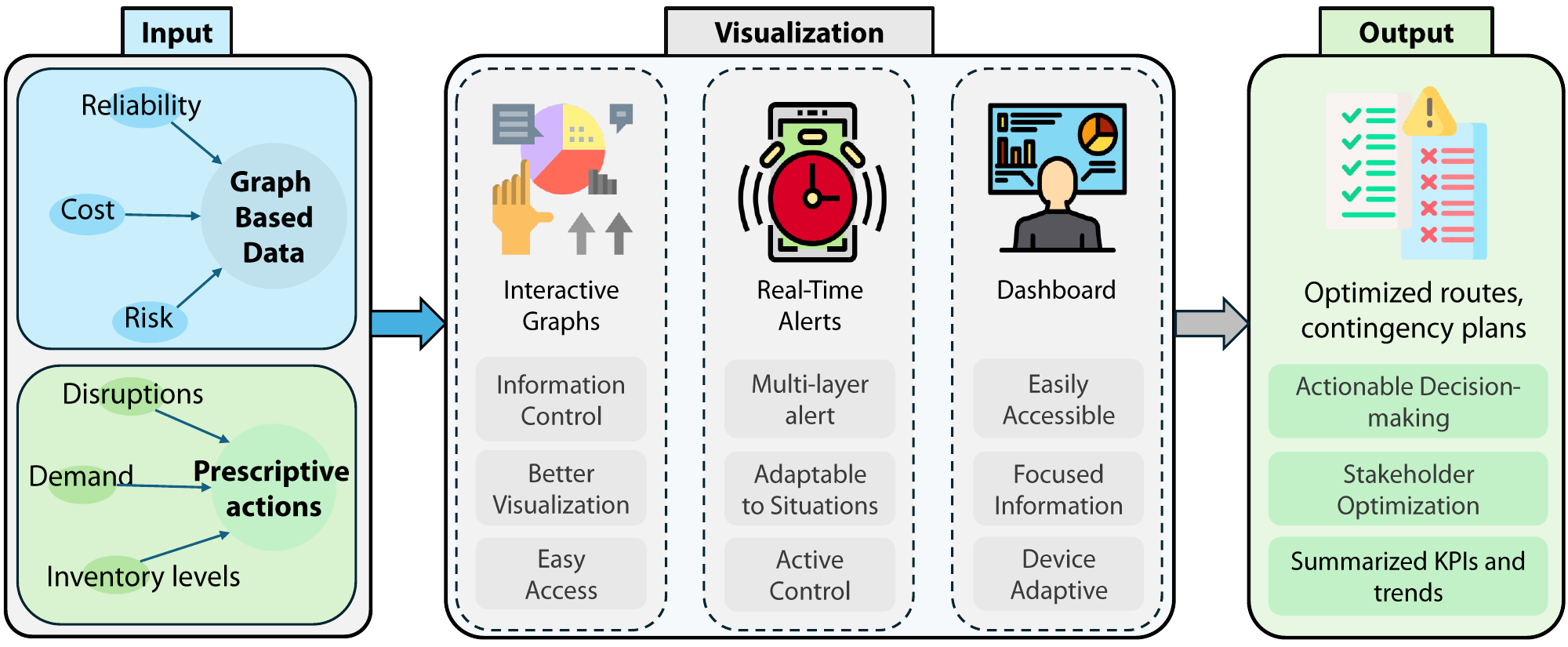}
    \caption{Workflow and Components of Visualization Interface}
    \label{fig:7_Visualization_Interface_Function}
\end{figure}

The Visualization Interface Function is the final layer of our Graph-based Digital Twin (DT) architecture, acting as the key bridge between advanced computational insights and practical decision-making. Building on the integrated data, structured graph models, and advanced simulations from the earlier layers, this stage transforms complex supply chain information into an interactive and intuitive interface, as demostrated in Figure \ref{fig:7_Visualization_Interface_Function}. This real-time visualization system provides stakeholders with a clear, dynamic overview of supply chain operations, highlighting key metrics, dependencies, and potential disruptions in an easily understandable format. More than just a monitoring tool, it empowers decision-makers by offering optimized routing suggestions, identifying vulnerability hotspots, and recommending contingency strategies. By integrating visual analytics with interactive features, we ensure that users can explore, analyze, and respond to supply chain fluctuations in real time. This capability is essential for handling modern supply chain complexities, fostering collaboration among teams, and promoting data-driven, proactive decision-making.

At the core of this layer are interactive graph visualizations that dynamically display supply chain elements. The nodes represent key entities—such as suppliers, manufacturers, warehouses, and customers—and provide real-time operational insights like inventory levels and production capacity \citep{prem_prakash_mishra__2017}. The edges illustrate relationships such as transportation routes, contractual agreements, and financial transactions, annotated with critical details like lead times, transportation costs, and reliability scores. For instance, an edge linking a warehouse to a distributor might indicate both the cost of transportation and the estimated delivery time, offering immediate visibility into bottlenecks and inefficiencies \citep{peng_ji__2024}. Additionally, a real-time alert system integrates insights from previous layers, instantly flagging disruptions like transportation delays or vulnerabilities in high-risk nodes. For example, if a port delay is detected through stress-testing algorithms in the Simulation and Analysis Engine, an alert would notify stakeholders, allowing them to review the issue and implement contingency plans directly from the interface \citep{claudia_pani__2013}. This ensures a proactive response to disruptions, helping to mitigate cascading failures before they affect broader supply chain operations.

Our dashboard component serves as the central hub for monitoring Key Performance Indicators (KPIs) that reflect the overall health and efficiency of the supply chain, enabling high-level strategic decision-making. By aggregating real-time data from earlier layers, the dashboard provides stakeholders with a clear, actionable overview of costs, time metrics, inventory levels, and sustainability factors. Cost metrics break down expenditures across supply chain nodes and edges, identifying inefficiencies that may inflate logistics or production costs. For instance, if certain transportation routes exhibit consistently high costs, stakeholders can analyze and optimize logistics strategies using shortest-path solutions derived from Layer 3 \citep{dedy_rahman_wijaya__2023}. Similarly, time metrics focus on lead times and delivery schedules, helping managers track delays in real time and adjust workflows accordingly. By integrating time-sensitive data from Layer 2's graph construction and Layer 3’s dynamic simulations, we ensure that decision-makers have reliable insights to improve operational efficiency. Additionally, inventory levels are continuously updated using real-time data feeds from Layer 1, preventing overstocking or stockouts and aligning supply with predictive demand models. Another critical component is carbon emissions tracking, which has become essential in sustainable supply chain management. By monitoring emissions from transportation and production activities, organizations can assess their environmental footprint and ensure compliance with regulatory standards \citep{blessing_ameh_2024}. These diverse KPIs are presented through intuitive visual formats such as heatmaps, trend graphs, and performance dashboards, allowing stakeholders to quickly interpret trends, identify problem areas, and take proactive measures. Ultimately, this dashboard not only enhances operational decision-making but also aligns business objectives with sustainability goals, ensuring a resilient, cost-effective, and environmentally responsible supply chain.

\begin{table}[ht]
\centering
\caption{Tools for Supply Chain Visualization in Digital Twins}
\label{tab:visualization_tools}
\begin{tabular}{|p{2cm}|p{3cm}|p{5cm}|p{4cm}|}
\hline
\textbf{Tool}           & \textbf{Specialization}                      & \textbf{Digital Twin Application}                                   & \textbf{Example Use Case}                         \\ \hline
Gephi                  & Large-scale graph rendering and clustering.  & Identifying supply chain bottlenecks and hidden dependencies.       & Clustering high-risk supplier nodes.             \\ \hline
Cytoscape.js           & Real-time, lightweight graph visualizations. & Integrates with dashboards to visualize real-time metrics.          & Interactive updates for inventory levels.         \\ \hline
D3.js                  & Customizable visualizations (heatmaps, flows). & Provides dynamic analysis of environmental impact or material flows. & Creating heatmaps for carbon emissions.          \\ \hline
Power BI               & KPI dashboards and predictive analytics.     & Enables users to explore cost, time, and inventory trends.          & Tracking delivery schedules and lead times.       \\ \hline
Neo4j Bloom            & Narrative-driven graph exploration.          & Simplifies complex data for stakeholder presentations.              & Highlighting critical distribution hubs.          \\ \hline
Kepler.gl              & Geospatial visualizations.                   & Monitors transportation routes and delivery statuses.               & Real-time tracking of shipments on interactive maps. \\ \hline
\end{tabular}
\end{table}

\subsubsection{Interactive Visualization Tools for Supply Chain Optimization}
Interactive visualization tools play a crucial role in transforming complex supply chain data into clear, actionable insights, allowing stakeholders to analyze, interpret, and optimize operations with confidence. These tools bridge the gap between raw data and strategic decision-making by offering intuitive, interactive environments that enable users to explore supply chain networks, identify hidden patterns, and assess key performance metrics. Graph visualization libraries provide dynamic representations of supply chain graphs, allowing us to track disruptions, detect inefficiencies, and enhance resilience \citep{kosasih2024towards}. For instance, Gephi, known for its scalability and 3D rendering, enables real-time network analysis, helping us identify supplier disruptions or pinpoint critical bottlenecks \citep{sbastien_heymann__2013}. With its clustering algorithms and modular design, it uncovers hidden dependencies within supply chains, supporting data-driven risk mitigation strategies. Similarly, Cytoscape.js, a lightweight JavaScript library, seamlessly integrates with real-time dashboards, allowing decision-makers to interact with inventory levels, transportation costs, and network flows \citep{max_franz__2023}. Additionally, D3.js enables customized visualizations, such as heatmaps for carbon emissions or flow diagrams for material movements, ensuring a more comprehensive understanding of supply chain efficiency \citep{p_bhat__2021}.

Beyond specialized graph visualization libraries, user-friendly interfaces such as Power BI and Tableau democratize data access and analysis, making advanced supply chain insights available to a wider range of users. Power BI enhances predictive analytics, enabling real-time monitoring of KPIs like costs, lead times, and inventory trends, ensuring organizations can respond proactively to fluctuating market conditions \citep{krishna_k__krishnan__2006}. Meanwhile, Tableau’s drag-and-drop interface allows stakeholders to visualize regional bottlenecks, compare shipping costs, and develop interactive dashboards, improving collaboration and strategic alignment \citep{jun_kim__2015}. Its ability to integrate with visualization libraries like D3.js further strengthens its capability to represent complex supply chain structures effectively. Emerging tools such as Neo4j Bloom and Kepler.gl further enrich visualization possibilities—Neo4j Bloom simplifies graph data exploration with narrative-driven interfaces, while Kepler.gl provides geospatial visualizations that track real-time transportation routes and distribution hubs \citep{rajesh_tamilmani__2019}. By leveraging these advanced visualization tools, we can navigate supply chain complexities more effectively, improve decision-making, and drive resilience across operations.

\subsubsection{Feedback Loop}
The feedback loop is an essential element of the Digital Twin (DT) framework, ensuring that the predictions and insights generated by the digital model remain aligned with real-world outcomes. This continuous cycle of refinement and validation enhances the accuracy and reliability of the DT, making it a powerful tool for data-driven decision-making. By incorporating real-time data and comparing it with simulated outcomes, we create a dynamic system that adapts to operational changes and improves over time. A well-designed feedback mechanism strengthens the DT’s ability to anticipate bottlenecks, inefficiencies, and risks, allowing organizations to proactively optimize their supply chain processes. The integration of a verification and validation (V\&V) framework ensures that the digital twin is both structurally sound and representative of reality, reinforcing its role as a trustworthy decision-support tool.

Verification and validation (V\&V) are fundamental to establishing the credibility of the digital twin framework. Verification ensures that the DT is built correctly, meaning its graph representations, simulation models, and algorithms function as designed without errors \citep{beatriz_cabrerodaniel__2024}. Validation, on the other hand, confirms that the DT accurately represents the real-world system, ensuring that its predictions and optimizations align with actual operational performance. To strengthen these processes, we can leverage modular ontologies, which provide structured approaches to integrating proprietary and third-party digital twins. Additionally, advanced models such as the quintuple helix framework, which incorporates perspectives from academia, industry, government, civil society, and environmental sustainability, enhance validation by ensuring that the DT reflects diverse datasets and real-world complexities \citep{ana_perii__2024}. By continuously refining the digital twin using iterative feedback, we ensure that it remains a robust, adaptive, and actionable tool for optimizing supply chain operations and strategic decision-making.

\begin{table}[ht]
\centering
\caption{Feedback Loop in Digital Twin Framework}
\label{tab:feedback_loop}
\begin{tabular}{|p{2.6cm}|p{3cm}|p{3.5cm}|p{4cm}|}
\hline
\textbf{Component}      & \textbf{Functionality}                       & \textbf{Digital Twin Role}                                   & \textbf{Example Improvement}                 \\ \hline
Verification           & Ensures Digital Twin functions per specifications. & Validates graph structures, simulations, and algorithms.     & Confirming accuracy of route optimization models. \\ \hline
Validation             & Assesses DT predictions against real-world outcomes. & Refines models for improved decision-making and accuracy.   & Adjusting forecasts based on supplier delivery data. \\ \hline
Continuous Updates     & Integrates real-time data to enhance model accuracy. & Keeps Digital Twin relevant to evolving supply chain conditions. & Recalibrating for seasonal demand fluctuations. \\ \hline
Causal Falsification   & Identifies prediction discrepancies.           & Ensures robust and reliable Digital Twin insights.          & Refining edge weights for updated transportation times. \\ \hline
\end{tabular}
\end{table}

Continuous validation is a critical aspect of the feedback loop, ensuring that the digital twin (DT) remains relevant and accurate as real-world conditions change over time. As new data streams in from sources such as IoT devices, ERP systems, and logistics databases, the DT continuously updates to reflect shifts in demand patterns, transportation routes, and supplier performance. This iterative validation process helps the model stay in sync with actual operational changes. For example, if the DT simulates a delay due to weather disruptions, real-time validation ensures that the predicted delay aligns with the actual impact observed \citep{jiwon_kim__2014}. Whenever discrepancies arise between the predicted and actual outcomes, adjustments are made to the model, improving its precision and adaptability. By regularly updating the model with real-world data, the DT becomes a more reliable tool for decision-making in dynamic environments.
A particularly promising approach to improving the DT’s accuracy is causal falsification, which utilizes causal inference techniques to identify potential errors in the predictions made by the model \citep{robert_cornish__2023}. Instead of assuming the model is always correct, this method actively searches for discrepancies between expected and observed results. For instance, in a supply chain scenario, if the DT predicts delays in a supplier’s delivery but the actual delays differ, causal falsification would highlight this difference \citep{bo_shi__2024}. In response, the model’s parameters—such as those representing transportation times or supplier reliability—can be reassessed and updated. This approach ensures that the digital twin evolves and refines its accuracy over time, improving its ability to represent the real world.

The continuous validation within the feedback loop also supports ongoing improvements in both prediction accuracy and the quality of the insights derived from the DT. For example, if inventory forecasts deviate from actual stock levels, we can adjust the inventory management model to enhance future predictions \citep{cisse_sory_ibrahima__2021}. Similarly, if transportation simulations don’t align with real disruptions, we can refine the route optimization algorithms to improve future performance \citep{hernan_chavez__2017}. This process of ongoing refinement not only boosts the predictive capabilities of the DT but also builds confidence among stakeholders, enabling them to make more informed decisions with greater certainty. By continuously iterating and improving the model, we ensure the DT’s effectiveness as a data-driven tool for proactive decision-making in complex supply chains.

\subsubsection{Process}
Our feedback loop layer plays a crucial role in ensuring the continuous improvement of the Digital Twin (DT) by integrating real-world data and updating its models for better accuracy and reliability. This iterative process allows the DT to remain relevant and effective over time \citep{mariam_abed__2023}. The process begins by monitoring the outcomes of DT-generated recommendations—such as optimized transportation routes, inventory management strategies, or supplier risk assessments—to evaluate their real-world effectiveness. For example, stakeholders track key operational metrics, including delivery times, fuel consumption, logistics costs, inventory turnover, stock-out rates, and supplier reliability, to assess whether the DT’s decisions, such as route optimization or stock adjustments, successfully reduced inefficiencies. If discrepancies emerge between predicted and actual outcomes, we analyze the underlying factors to pinpoint gaps, missed predictions, or partial successes \citep{hernan_chavez__2017}. For instance, even if the DT optimizes a delivery route, unforeseen events like weather disruptions or road closures may still cause delays, requiring further refinements in the model. Similarly, an inventory adjustment might prevent shortages in one warehouse while unintentionally leading to overstock in another \citep{yingbin_zhang__2023}. These insights serve as critical inputs for enhancing the DT’s predictive capabilities.

To refine the DT, we update its graph structures, machine learning models, and reinforcement learning algorithms with new data. For example, a Graph Neural Network (GNN) that forecasts demand fluctuations may adjust its weight parameters based on updated sales trends \citep{wasi}. Likewise, a Deep Q-Network (DQN) designed for inventory management may recalibrate its stock recommendations to prevent future stock-outs \citep{oroojlooyjadid2022deep}. Additionally, we update the graph representation of the supply chain, modifying node attributes (e.g., supplier reliability, warehouse storage capacity) and edge weights (e.g., transportation delays, route efficiency) to reflect recent changes in real-world operations. This refined DT then generates improved recommendations, such as better transportation routes that account for new disruptions, enhanced supplier risk assessments, and more precise inventory predictions that minimize overstocking or shortages.

Impact of an improved feedback loop is visible in many tangible business benefits. For instance, enhanced demand forecasting using DT-driven insights has been shown to reduce stock-outs by 22\% \citep{monica_balderas__2019}, while logistics cost optimizations through alternative routing strategies have led to 15\% savings in expenses \citep{jose_l__marzo__2007}. Additionally, by integrating resilience measures, such as identifying high-risk nodes and edges in the supply chain, companies can proactively mitigate disruptions and strengthen overall supply chain robustness \citep{krishna_hajarath__2024}. By seamlessly incorporating real-time feedback into its decision-making models, the DT evolves into a highly adaptive tool, empowering organizations to navigate complex and ever-changing supply chain environments with confidence.

\begin{table}[ht]
\centering
\caption{Dynamic Refinement of Digital Twin Systems}
\label{tab:feedback_process}
\begin{tabular}{|p{3cm}|p{4cm}|p{4cm}|p{4cm}|}
\hline
\textbf{Step}          & \textbf{Action}                                & \textbf{Digital Twin Adaptation}                           & \textbf{Outcome}                          \\ \hline
Data Collection       & Collects outcomes of DT-recommended actions.   & Evaluates predictions and updates models.                 & Improved inventory turnover rates.         \\ \hline
Gap Analysis          & Identifies discrepancies in predictions.       & Refines simulation parameters and model attributes.       & Enhanced accuracy of demand forecasts.     \\ \hline
Model Refinement      & Updates simulations and algorithms.            & Integrates new data into DT for better predictive accuracy. & Reduced logistics costs via optimized routes. \\ \hline
Recalibrated Insights & Generates updated insights and recommendations. & Ensures actionable, relevant outputs for decision-makers.  & Faster response to supply chain disruptions. \\ \hline
\end{tabular}
\end{table}

\section{Applications and Case Studies}

\begin{table}[h!]
\caption{Industry Problems and Solutions with GDT Potential}
\centering
\scalebox{0.65}{
\begin{tabular}{|p{3.5cm}|p{4cm}|p{2.5cm}|p{3.5cm}|p{5cm}|p{5cm}|}
\toprule
\textbf{Industry Name} & \textbf{Specific Problem} & \textbf{Category} & \textbf{Current Solution Used} & \textbf{Benefits of Current Solution} & \textbf{Further Benefits with GDT} \\ 
\midrule
Food and Beverage & High spoilage rates in perishable goods & Sustainability & IoT sensors to monitor temperature & Reduced waste and improved product quality & Dynamic visualization of spoilage risks and optimized storage strategies \\ \hline
Retail & Stockouts and overstocking in inventory & Optimization & Predictive analytics for demand forecasting & Better inventory management and reduced costs & Real-time adjustment of inventory levels with better network visibility \\  \hline
Pharmaceuticals & Non-compliance with temperature regulations & Real-time Monitoring & Environmental monitoring sensors & Enhanced regulatory compliance and safety & Enhanced tracking of compliance across supply chain nodes \\  \hline
Logistics & Delays in delivery during peak seasons & Real-time Monitoring & AI-powered route optimization tools & Improved delivery times and customer satisfaction & Proactive rerouting and real-time disruption management \\  \hline
Automotive & Disruption in supply due to single-sourcing & Resilience Analysis & Supplier diversification strategies & Reduced risk of supply chain disruptions & Holistic risk analysis and adaptive response strategies \\  \hline
Textile & Excessive water usage in production & Sustainability & Water recycling systems in factories & Lower water consumption and reduced costs & Better monitoring and simulation of resource consumption \\  \hline
E-commerce & Inefficient last-mile delivery & Optimization & AI-based delivery route optimization & Reduced delivery times and operational costs & Enhanced coordination for on-the-go delivery changes \\  \hline
Electronics & Frequent component shortages & Resilience Analysis & Inventory buffer strategies & Minimized production delays & Dynamic modeling of supply chain dependencies \\  \hline
Energy & Variability in renewable energy supply & Real-time Monitoring & Smart grid systems with real-time monitoring & Improved energy stability and usage efficiency & Improved demand-supply matching across energy networks \\  \hline
Healthcare & Delays in medical supply restocking & Resilience Analysis & Automated inventory management systems & Faster replenishment and improved patient care & Real-time visualization of medical supply chain dependencies \\ 
\bottomrule
\end{tabular}}
\label{tab:industry_problems}
\end{table}

\subsection{Resilience Analysis}
Supply chain resilience refers to the ability to anticipate, adapt to, and recover from disruptions while maintaining smooth operations. Key vulnerabilities include single-supplier dependence, lean inventory practices, and global uncertainties like natural disasters, pandemics, and geopolitical instabilities. Events such as the COVID-19 pandemic and the 2011 Tohoku earthquake revealed how disruptions at critical points can trigger cascading failures \cite{tinglong_dai__2024}. To enhance resilience, we must adopt proactive and reactive strategies. Proactive approaches involve diversifying suppliers, maintaining buffer stocks, and leveraging predictive analytics, such as Failure Mode and Effects Analysis (FMEA) \cite{chen2013modified} and Monte Carlo simulations \cite{belvardi2012monte}, to identify risks early. Reactive measures, like dynamic routing algorithms, digital control towers, and scenario planning tools such as AnyLogic\footnote{https://www.anylogic.com/} and Simul8\footnote{https://www.simul8.com/}, help businesses respond swiftly to disruptions. Blockchain technology further enhances resilience by improving traceability and transparency. Case studies, such as Toyota’s response to the 2011 tsunami, highlight the effectiveness of combining supplier redundancy with agile response mechanisms \cite{frans_lavdari__2024}. A Graph-based Digital Twin (GDT) offers a powerful solution by creating a real-time digital replica of the supply chain, enabling predictive analysis, scenario simulations, and rapid decision-making. During the COVID-19 pandemic, firms using GDTs successfully analyzed supply chain interdependencies and adapted to shifting demand patterns \cite{lu_wang__2022}. By integrating machine learning and graph analytics, GDTs improve resilience, agility, and efficiency, ensuring a robust, future-ready supply chain capable of navigating unforeseen challenges.

\subsection{Supply Chain Optimization}
Supply chain optimization aims to reduce costs, improve efficiency, and meet customer needs, but challenges like fluctuating demand, supply disruptions, and real-time decision-making make this difficult. Complex interdependencies in global networks lead to issues such as the bullwhip effect, inefficient resource allocation, and long lead times \cite{melek_akn_ate__2023}. Balancing cost reduction with operational flexibility requires advanced strategies. Machine learning methods, like neural networks and support vector machines, enhance demand forecasting, while VRP and TSP algorithms improve route optimization by adapting to real-time traffic conditions \cite{lszl_kovcs__2024}. Simulation-optimization models help manage uncertainties, as seen in manufacturing, where they optimize lead time, risk, and cost. Dynamic inventory management powered by predictive analytics minimizes overstocking and stockouts \cite{kamal_ola_alamin__2024}. Graph-based Digital Twins (GDTs) take optimization further by transforming static models into dynamic, interconnected systems, using real-time data and advanced analytics to uncover inefficiencies. By simulating different strategies, GDTs refine demand forecasting, enhance route planning, and improve resource allocation. A logistics firm, for example, used graph-based AI-driven routing algorithms to optimize path planning, reducing travel time and costs by adjusting routes dynamically \cite{ruiping_li__2024}. With these capabilities, GDTs introduce precision and agility, making supply chain operations more resilient and efficient.

\subsection{Supply Chain Sustainability}
Supply chain sustainability focuses on managing environmental, social, and economic impacts across a product’s lifecycle, ensuring responsible governance while balancing profitability with ethical and environmental concerns. Challenges arise due to the complexity of global supply chains, where limited transparency and accountability often lead to resource exploitation and unethical labor practices \cite{hibanan_adiid_2023}. The lack of standardized regulations and resistance to cost increases further hinder sustainable efforts. To address these challenges, companies use Lifecycle Assessment (LCA) tools to evaluate carbon footprints and identify areas for improvement \cite{kathleen_b__aviso_2024}. Circular economy principles, such as resource reuse and recycling, help minimize waste, while supplier engagement programs ensure compliance with sustainability standards. Many companies are also adopting renewable energy for manufacturing and greener logistics solutions, like electric vehicles, to reduce emissions \cite{gbor_nagy__2024}. Graph-based Digital Twins (GDTs) enhance these efforts by modeling carbon footprints in real time, allowing companies to pinpoint high-emission areas and optimize resource allocation. For example, GDT simulations of greener transportation routes have successfully reduced fuel costs and emissions \cite{kbra_duran__2024}. By improving collaboration and transparency among suppliers, manufacturers, and distributors, GDTs enable a more adaptive and resilient approach to sustainability.

\section{Discussion} \label{sec:discussion}

Our study highlights the growing challenges in supply chain management, emphasizing how traditional systems struggle with fragmented data, inefficiencies, and a lack of sustainability considerations. By leveraging a Graph-Based Digital Twin (GDT) framework, we offer a structured approach that enables real-time monitoring, optimization, and predictive analytics for supply chain networks. The integration of graph modeling with DTs allows businesses to uncover hidden inefficiencies, dynamically adjust to market fluctuations, and enhance collaboration across stakeholders. Our proposed framework addresses key limitations in existing models by embedding sustainability metrics directly into operational dashboards, ensuring that decision-making aligns with environmental and resource efficiency goals \cite{hibanan_adiid_2023, kathleen_b__aviso_2024, kbra_duran__2024, ruiping_li__2024}. By bridging the gap between scalability and sustainability, we believe our approach fosters supply chains that are not only cost-effective but also resilient and environmentally responsible.

\textbf{Societal Impact.} \quad The adoption of GDT frameworks in supply chain optimization extends beyond operational efficiency; it has profound societal implications. Traditional supply chains often lack transparency, leading to unethical labor practices, unchecked resource exploitation, and supply-demand mismatches that exacerbate economic disparities. Our approach enhances visibility across supply networks, fostering accountability and ethical sourcing \cite{melek_akn_ate__2023, lsxl_kovacs__2024, kamal_ola_alamin__2024, gbor_nagy__2024}. Additionally, by incorporating sustainability metrics, businesses can reduce carbon footprints and contribute to climate action, aligning with global sustainability goals. The ability to predict and mitigate disruptions in real time also benefits communities by ensuring stable access to essential goods, particularly in times of crisis. Furthermore, by integrating renewable energy sources and optimizing transportation, our framework supports long-term environmental sustainability, directly benefiting both industry and society at large.

\textbf{Managerial Insights for Decision Makers in Industry.} \quad Implementing a GDT-based supply chain optimization model presents significant advantages for managers seeking to enhance agility, reduce costs, and improve strategic decision-making. By leveraging real-time data analytics and graph-based insights, managers can proactively identify bottlenecks, optimize resource allocation, and enhance operational resilience \cite{kbra_duran__2024, subrat_sahoo_2022, kavitha_kumari__k_s__2024, ruiping_li__2024}. The dynamic nature of our framework allows organizations to simulate various scenarios, assess potential risks, and develop contingency plans before disruptions occur. This proactive approach enables better supply chain coordination, improving overall responsiveness to market changes. Moreover, embedding sustainability metrics within dashboards not only aligns operations with regulatory requirements but also enhances corporate social responsibility (CSR) initiatives. Companies that embrace GDT-driven models gain a competitive edge by transitioning from reactive to predictive strategies, ensuring long-term efficiency, resilience, and sustainability in their supply chain operations.

\section{Limitations, Challenges and Future Research Directions} \label{sec:Challenges-and-research-directions}

\paragraph{Data Availability and Quality}
A major challenge in implementing graph-based digital twins (GDTs) for supply chain networks (SCNs) is ensuring access to high-quality, real-time data. Many organizations rely on disparate data sources, often lacking standardized formats, which leads to inconsistencies in information retrieval and integration. The challenge is further amplified when dealing with suppliers and partners across different geographic regions with varying levels of digital infrastructure. Additionally, real-time data collection often depends on IoT devices, RFID tags, and sensor networks, which may introduce noise, missing values, or latency issues. Without reliable data, the accuracy of predictive models and optimization strategies can be significantly compromised \cite{melek_akn_ate__2023}. Future research should focus on developing robust data harmonization frameworks that use machine learning techniques to clean, validate, and enhance data streams, ensuring that digital twins remain a precise representation of real-world supply chains.

\paragraph{Potential for Interdisciplinary Applications}
Integration of GDTs in supply chain management extends beyond logistics and operations, offering potential interdisciplinary applications. For instance, in healthcare, digital twins could model pharmaceutical supply chains to prevent drug shortages, while in manufacturing, they could optimize resource allocation for sustainable production \cite{kathleen_b__aviso_2024}. Other fields, such as environmental science, can leverage GDTs to monitor carbon footprints and enhance sustainability efforts by analyzing lifecycle impacts. Additionally, social sciences can use these models to understand labor conditions and ethical supply chain practices. Exploring these interdisciplinary avenues will help bridge knowledge gaps and drive innovations across sectors. Future studies should focus on collaborative research across disciplines, ensuring that GDT frameworks are adaptable and beneficial across multiple domains.

\paragraph{Role of Generative Models and Pre-Trained GNNs in SCNs}
Generative models and pre-trained Graph Neural Networks (GNNs) present an opportunity to improve SCN simulations by enhancing predictive accuracy and adaptability. Generative models, such as Variational Autoencoders (VAEs) and Generative Adversarial Networks (GANs), can be used to simulate demand fluctuations, supply disruptions, and market dynamics, helping businesses prepare for uncertainties \cite{ruiping_li__2024}. Similarly, pre-trained GNNs trained on large-scale supply chain data can be fine-tuned for specific industries, reducing computational costs and improving generalization across different networks. Despite their potential, challenges remain in training these models on diverse, high-quality datasets that accurately reflect the complexities of global supply chains. Future research should explore efficient pre-training strategies, domain adaptation techniques, and benchmark datasets tailored for SCN applications.

\paragraph{Scalability}
As supply chains grow increasingly complex, scalability becomes a crucial concern in implementing GDTs. The computational burden of processing vast networks with millions of nodes and edges can hinder real-time analysis. Traditional graph algorithms often struggle to handle such large-scale data, necessitating more efficient parallel computing and distributed processing methods \cite{kbra_duran__2024}. Additionally, scalability issues arise when integrating multiple data streams from various stakeholders, requiring robust architectures that balance accuracy and computational feasibility. Future research should investigate scalable graph-based ML techniques, such as graph sampling, hierarchical GNNs, and cloud-based processing solutions, to ensure that GDTs remain efficient even for large, multi-tiered supply networks.

\paragraph{Interoperability}
One of the biggest obstacles in adopting GDTs is the integration of diverse data sources and systems. Supply chains involve multiple stakeholders using different enterprise resource planning (ERP) software, warehouse management systems, and third-party logistics platforms, leading to interoperability challenges \cite{subrat_sahoo_2022}. Standardizing data exchange formats and establishing common communication protocols can help mitigate these issues. Moreover, advancements in API-based integrations and blockchain technology offer promising solutions for secure and transparent data sharing. Future research should focus on developing interoperable frameworks that seamlessly connect disparate systems while maintaining data integrity and security.

\paragraph{Ethical Considerations}
With increased data sharing and real-time monitoring in supply chains, ethical considerations such as privacy and security must be addressed. Sensitive business data, including supplier relationships and inventory levels, must be protected against cyber threats and unauthorized access \cite{kavitha_kumari__k_s__2024}. Additionally, data privacy concerns arise when monitoring employees, suppliers, and logistics partners, raising ethical questions about surveillance and consent. Implementing secure encryption methods, access control mechanisms, and compliance with global data protection regulations (e.g., GDPR) will be essential in mitigating these risks. Future research should focus on developing ethical AI frameworks and privacy-preserving techniques, such as federated learning, to ensure that supply chain optimization does not come at the expense of data security and individual rights.

\paragraph{Future Research Directions}
Looking ahead, future research should advance graph-based ML techniques, improve cross-domain digital twins, and explore the policy implications of GDT adoption. Novel approaches in dynamic graph modeling, reinforcement learning for supply chain decision-making, and self-learning digital twins could enhance adaptability and efficiency. Cross-domain applications of digital twins, integrating fields such as climate modeling and disaster resilience, offer promising areas for exploration \cite{gbor_nagy__2024}. Additionally, policymakers must consider regulations that promote transparency and sustainability while encouraging innovation in supply chain digitalization. Establishing global standards for digital twin adoption and ethical AI practices will be key in shaping the future of supply chain optimization. By addressing these research gaps, we can drive forward the development of intelligent, resilient, and sustainable supply networks.

\section{Conclusion} \label{sec:conclusion}
In this work, we proposed a Graph-Based Digital Twin (GDT) Framework for Supply Chain Optimization, addressing the limitations of traditional supply chain models in handling complex interdependencies, real-time monitoring, and sustainability concerns. Our approach integrates graph modeling to capture intricate relationships within supply networks and digital twin technology to enable dynamic simulations and real-time decision-making. The framework consists of a Data Integration Layer for harmonizing diverse data sources, a Graph Construction Module for modeling network dependencies, and a Simulation and Analysis Engine for optimizing operations. Additionally, we incorporated sustainability metrics, such as carbon footprints and resource utilization, into decision dashboards to support environmentally responsible supply chain management.

By leveraging the synergy between graph-based machine learning and digital twins, our framework enhances scalability, adaptability, and predictive capabilities in supply chain operations. Through real-time monitoring and data-driven simulations, businesses can proactively mitigate disruptions, optimize logistics, and improve resource allocation. We also highlighted the role of advanced machine learning techniques, such as pre-trained graph neural networks (GNNs) and generative models, in refining supply chain decision-making.

Our study contributes to both academia and industry by proposing a novel framework that balances efficiency, resilience, and sustainability in modern supply chains. Future research should explore cross-domain applications of GDTs, enhance interoperability across different data systems, and address the ethical challenges of real-time data sharing. As supply chains continue to evolve in complexity, our framework provides a scalable and adaptive foundation for next-generation intelligent supply chain management systems.

\clearpage
\newpage

\renewcommand{\bibfont}{\footnotesize} 
\setlength{\bibsep}{0pt} 
\bibliography{main}


\end{document}